\documentclass[12pt,a4paper]{article}
\usepackage{graphicx}
\usepackage{fancyhdr}
\usepackage{setspace}

\usepackage{url}
\DeclareUrlCommand\url{\def\UrlLeft{<}\def\UrlRight{>} \urlstyle{tt}}

\usepackage{xcolor}
\newcommand{\new}[1]{{\color{red}#1}}
\newcommand{\bfit}[1]{{\bf \textit{#1}}}
\usepackage{tabularx}

\newcommand{\txtsup}[1]{\ensuremath{^{\textrm #1}}}
\newcommand\vek[1]{\ensuremath{\mathbf{#1}}}

\newcommand{\strain}{\varepsilon}

\newcommand{\stress}{\sigma}
\newcommand{\tenss}[1]{\mbox{\boldmath$#1$}}

\begin{document}
\title{Soft computing-based calibration of microplane M4 model
parameters: Methodology and validation}
\author{A. Ku\v{c}erov\'{a}, M. Lep\v{s}\footnote{Corresponding author. Tel. +420 224 355 326; fax: +420 224 310 775. E-mail address: leps@cml.fsv.cvut.cz}}
\maketitle
\centerline{Department of Mechanics, Faculty of Civil Engineering}
\centerline{Czech Technical University in Prague}
\centerline{Th\'{a}kurova 7, 166 29 Prague, Czech Republic}

%%%%%%%%%%%%%%%%%%%%%%%%%%%%%%%%%%%%%%%%%%%%%%%%%%%%%%%%%%%%%%%%%%%%%
%%%%%%%%%%%%%%%%%%%%%%%%%%%%%%%%%%%%%%%%%%%%%%%%%%%%%%%%%%%%%%%%%%%%%
\vspace{0.2in}
\centerline{Dedicated to Professor Zden\v{e}k Bittnar in occasion of his 70th birthday.}
\vspace{0.2in}
%%%%%%%%%%%%%%%%%%%%%%%%%%%%%%%%%%%%%%%%%%%%%%%%%%%%%%%%%%%%%%%%%%%%%
%%%%%%%%%%%%%%%%%%%%%%%%%%%%%%%%%%%%%%%%%%%%%%%%%%%%%%%%%%%%%%%%%%%%%

%%%%%%%%%%%%%%%%%%%%%%%%%%%%%%%%%%%%%%%%%%%%%%%%%%%%%%%%%%%%%%%%%%%%%
%%%%%%%%%%%%%%%%%%%%%%%%%%%%%%%%%%%%%%%%%%%%%%%%%%%%%%%%%%%%%%%%%%%%%
\begin{abstract}
  Constitutive models for concrete based on the microplane concept
  have repeatedly proven their ability to well-reproduce non-linear
  response of concrete on material as well as structural scales. The
  major obstacle to a routine application of this class of models is,
  however, the calibration of microplane-related constants from
  macroscopic data. The goal of this paper is two-fold: (i)~to
  introduce the basic ingredients of a robust inverse procedure for
  the determination of dominant parameters of the M4 model proposed by
  Ba\v{z}ant and co-workers in~\cite{Bazant:2000:MM4I} based on
  cascade Artificial Neural Networks trained by Evolutionary Algorithm
  and (ii)~to validate the proposed methodology against a
  representative set of experimental data. The obtained results
  demonstrate that the soft computing-based method is capable of
  delivering the searched response with an accuracy comparable to the
  values obtained by expert users.
\end{abstract}
%%%%%%%%%%%%%%%%%%%%%%%%%%%%%%%%%%%%%%%%%%%%%%%%%%%%%%%%%%%%%%%%%%%%%
%%%%%%%%%%%%%%%%%%%%%%%%%%%%%%%%%%%%%%%%%%%%%%%%%%%%%%%%%%%%%%%%%%%%%
%%%%%%%%%%%%%%%%%%%%%%%%%%%%%%%%%%%%%%%%%%%%%%%%%%%%%%%%%%%%%%%%%%%%%
%%%%%%%%%%%%%%%%%%%%%%%%%%%%%%%%%%%%%%%%%%%%%%%%%%%%%%%%%%%%%%%%%%%%%

\newpage
%%%%%%%%%%%%%%%%%%%%%%%%%%%%%%%%%%%%%%%%%%%%%%%%%%%%%%%%%%%%%%
%%%%%%%%%%%%%%%%%%%%%%%%%%%%%%%%%%%%%%%%%%%%%%%%%%%%%%%%%%%%%%
%%%%%%%%%%%%%%%%%%%%%%%%%%%%%%%%%%%%%%%%%%%%%%%%%%%%%%%%%%%%%%
%%%%%%%%%%%%%%%%%%%%%%%%%%%%%%%%%%%%%%%%%%%%%%%%%%%%%%%%%%%%%%
\section{Introduction}
\label{sec:introduction}

Despite long-term use of concrete in civil engineering industry,
development of an exhaustive constitutive model for concrete still
remains in the focus of engineering materials science. The difficulty
in constitutive modeling is directly linked to the complexity of the
material itself: the quasi-brittle response of concrete as seen on
macroscale is a consequence of non-linear interactions of its numerous
constituents over a wide range of lengthscales. As a result, a
formulation of a~concise model reproducing experimental data while
satisfying the constraints of continuum thermodynamics is far from
being trivial.

One of the most successful modeling approaches stems from the
microplane concept, in which the response of a macroscopic material
point results from contributions of planes of all possible
orientations, see e.g.~\cite[Chapter 25]{Jirasek:2001:IAS}. The
potential of this modeling paradigm was perhaps best demonstrated by
the M4 model for concrete introduced by Ba\v{z}ant and
co-workers~\cite{Bazant:2000:MM4I}, which has repeatedly shown its
capacity to realistically reproduce response of complex
three-dimensional structures under general loading conditions. In
addition, an efficient parallelization strategy~\cite{Nemecek:2002:MM}
or model adaptivity~\cite{Cervenka:2005:ELE} were proposed to
compensate for an increased computational cost of the microplane
models with respect to traditional approaches.

The major strength of the microplane concept -- the implicit format of
the macroscopic constitutive law -- is however closely linked to its
major obstacle: the model constants are directly linked to the
microplane level and as such are difficult to interpret and identify
from experiments performed at the macroscopic
level. In the particular case of
the M4 model for concrete, Caner and Ba\v{z}ant proposed in~\cite{Caner:2000:MM4II} a
heuristic sequence of experiments to calibrate adjustable parameters
of the model. Nevertheless, the procedure is still based on a
hand-fitting procedure, thereby requiring an expert user or a
tedious trial-and-error calibration procedure. To relax this
constraint, hereafter we introduce a robust automated
procedure for the calibration of the M4 model constants.

An independent certificate to the non-trivial character of the
automated M4 model calibration is provided by difficulties associated
with the selection of an appropriate class of optimization algorithms
to solve the resulting optimization problem based on the minimization
of the objective function. Following the well-established
classification of the inverse algorithms introduced by Mahnken
in~\cite{Mahnken:2004:ECM}, the most efficient solution strategies are
typically based on gradient-based optimization methods. Although
several successful applications to quasi-brittle materials have been
reported in the past, e.g.~\cite{Mahnken:1996:PIVP}, they cannot be
applied in the microplane setting since a~closed-form tangent is not
available, resulting in a~non-smooth objective function.

The requirements of smoothness can be relaxed by resorting to the
evolutionary algorithms. Although these methods were successfully
applied to inverse analysis of a variety of engineering constitutive
models, see e.g.~\cite{Kucerova:2007:PHD} and references therein, their extremely high demand for a number of function calls combined with computational cost of the microplane model prohibits their use even
in the parallel environment, see~\cite{Leps09:157924}.

To reduce the high computational cost of the resulting problem, one
may reconcile oneself to an approximate solution by resorting to an
inexpensive approximation of the quantity of interest. At this step,
network-type approximations are frequently employed due to their
simplicity and extremely high degree of flexibility. In general, two
complementary strategies to an approximate identification problem can
be distinguished~\cite{Kucerova:2007:PHD}: (i)~the forward mode, where
either the systems' response or the objective function itself is to be
replaced and (ii)~the inverse mode where an approximate map from
measurable quantities to model parameters is to be established. The
former model was, e.g., successfully applied to complex geotechnical
problems~\cite{Pichler:2003:BAMP} or to an identification of
parameters for damage-plasticity models from large-scale tests by
Ku\v{c}erov\'{a} et. al~\cite{Kucerova:2009}. Extensions towards the
microplane case, however, suffer from the fact that an accurate
representation of the load-displacement curve or the objective
function itself still requires a significant number of degrees of
freedom, which, when to be determined reliably, leads to an excessive
number of simulations. Moreover, a high accuracy of approximations is
usually needed to obtain the identified parameters with an acceptable
accuracy~\cite{Kucerova:2009}. Therefore, high computational costs
make it difficult to successfully implement this strategy for the
microplane identification problem.

The inverse identification mode, on the other hand, is substantially
more computationally efficient \new{and} is being widely used in a
wide range of material calibration problems, see
again~\cite{Kucerova:2007:PHD} and references therein. The critical
issue in this case remains the inherent ill-conditioning of the
problem, especially when determining a parameter with a low
sensitivity for the executed experiment. This fact was exactly the
reason for the failure of our early applications of the inverse mode
to the microplane identification problem, see~\cite{Drchal:2002:ECT}
for a particular illustration of this issue. Fortunately, efficient
small-size procedures are currently available to estimate the
sensitivity of a particular parameter and, as demonstrated by our
recent work~\cite{Kucerova10:176025}, present a natural choice to
identify the model parameters.

In the current work, we introduce a well-defined procedure for the
identification of material parameters for M4 microplane model. The
procedure itself is based on work by Nov\'ak and Lehk\'y \cite{Novak:2006:EAAN} and has been
successfully verified against artificially generated sets of data
in~\cite{Kucerova:2007:BAMM}. Here, we present an independent validation and comparison
with expert user guesses.

The remainder of the paper is organized as follows. The next section
is devoted to the short introduction and a description of our
implementation of M4 microplane model. The inverse mode of parameters'
estimation is presented next with an emphasis put on artificial neural
networks and their training. The main part is describing the
methodology behind the identification of all pertinent M4 model
parameters. To get reliable results, two ``tricks'' are used. The
first one is usage of cascade neural networks~\cite{Waszczyszyn:2005}
where some already identified parameters serve as inputs to later
stages of identification. The second one is the combination of two
models for closely interacting parameters. \new{The} combination
defines \new{a} nonlinear equation which \new{can be} easily
\new{solved numerically}. The main part is -- according to M4 authors
-- divided to three sections \new{corresponding to the} tests needed
for M4 parameters estimation: the uniaxial, hydrostatic and triaxial
tests. In the closing section, the presented text is
accompanied by the analysis of the computational demands which present
the main drawbacks of the presented methodology.

\section{COMPUTATIONAL MODEL}
%%%%%%%%%%%%%%%%%%%%%%%%%%%%%

\begin{figure}[th]
\centering
\includegraphics[width=5cm,keepaspectratio]{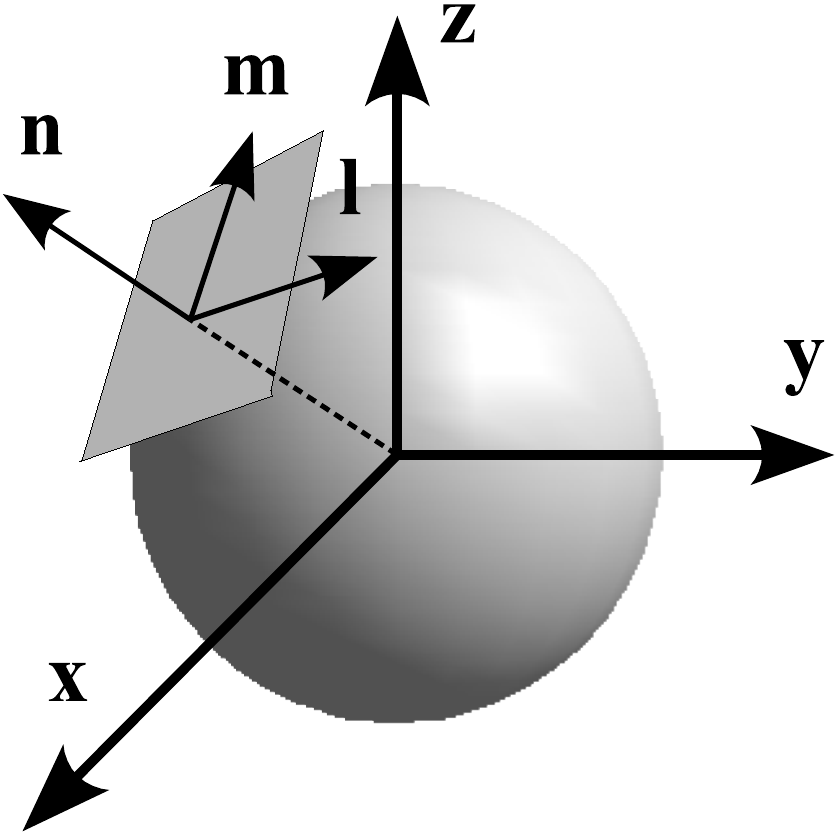}
\caption{Concept of microplane modeling} \label{fig:microplane}
\end{figure}

In contrary to traditional approaches to constitutive modeling, which
build on description via second-order strain and stress {\em tensors}
at individual points in the $( x, y, z )$ coordinate system, the
microplane approach builds the descriptions on planes of arbitrary
spatial orientations -- so-called {\em microplanes}, related to
a~macroscopic point, see Figure~\ref{fig:microplane}.  This allows to
formulate constitutive equations in terms of stress and strain {\em
  vectors} in the coordinate system $( \vek{l}, \vek{m}, \vek{n} )$
associated with a~microplane oriented by a~normal vector $\vek{n}$.
The general procedure of evaluation of a~strain-driven microplane
model response for a~given ``macroscopic'' strain tensor
$\tenss{\strain}( \vek{x} )$ can be described as follows: (i)~for a
given microplane orientation $\vek{n}$ normal ``macroscopic'' strain
tensor $\tenss{\strain}( \vek{x} )$ is projected onto the normal
``microstrain'' vector $\vek{\strain}( \vek{n})$ and the shear
microstrains $\vek{\strain}( \vek{m} )$ and $\vek{\strain}( \vek{l}
)$, (ii)~the normal and shear microstresses $\vek{\stress}( \vek{n} ),
\vek{\stress}( \vek{m} )$ and $\vek{\stress}( \vek{l} )$ are evaluated
using microplane constitutive relations, (iii)~the ``macroscopic''
stress tensor $\tenss{\stress}( \vek{x} )$ is reconstructed from the
microscopic ones using the principle of virtual work, see,
e.g.,~\cite[Chapter~25]{Jirasek:2001:IAS} for more details. In the
particular implementation, 28 microplanes with a~pre-defined
orientation on the unit hemisphere \new{are} used to evaluate the
response of the model.

To close the microplane model description, the appropriate microplane
constitutive relation must be provided to realistically describe
material behavior. The model examined in the current work is the
microplane model M4~\cite{Bazant:2000:MM4I}. The model uses
volumetric-deviatoric split of the normal components of the stress
and strain vectors, treats independently shear components of a
microplane and introduces the concept of ``boundary curves'' to limit
unrealistically high values predicted by earlier version of the
model. As a~result, the strain-to-stress map $\tenss{\strain}( \vek{x}
) \mapsto \tenss{\stress}( \vek{x} )$ is no longer smooth, which
complicates the formulation of consistent tangent stiffness
matrix~\cite{Nemecek:2002:MM} and, subsequently, gradient-based
approaches to material model parameters identification.

In overall, the microplane model~M4 needs seven parameters to describe
a~certain type of~concrete, namely: Young's modulus $E$, Poisson's
ratio $\nu$, and other five parameters ($k_1$, $k_2$, $k_3$, $k_4$,
$c_{20}$)\footnote{The addition of the constants $c_{18}$ -- $c_{20}$
  into M4 formulation and addition of $c_{20}$ into the parameters'
  set is discussed in~\cite{Nemecek00:73353}. The inclusion of
  $c_{20}$ is also confirmed by the validation phase in
  Section~\ref{sec:uni} of this contribution.}, which do not have
a~simple physical interpretation, and therefore it is difficult to
determine their values from experiments. The only information
available in the open literature are the bounds shown in the
Table~\ref{tbl1}.

\begin{table}[ht]
\centering
\begin{tabular}{rcl}
\hline
{\bf Parameter} & & {\bf Bounds} \\
\hline
$E$      & $\in$ & $\langle 20.0, 50.0 \rangle$ \mbox{ GPa} \\
$\nu$    & $\in$ & $\langle 0.1, 0.3 \rangle$   \\
$k_1$    & $\in$ & $\langle 0.00008, 0.00025 \rangle$   \\
$k_2$    & $\in$ & $\langle 100.0, 1000.0 \rangle$ \\
$k_3$    & $\in$ & $\langle 5.0, 15.0 \rangle$ \\
$k_4$    & $\in$ & $\langle 30.0, 200.0 \rangle$ \\
$c_{20}$ & $\in$ & $\langle 0.2, 5.0 \rangle$ \\
\hline
\end{tabular}
\caption{Bounds for the microplane model parameters}
\label{tbl1}
\end{table}

In the present work, the computational model of a~structure is
provided by the object-oriented C++ finite element code
OOFEM~1.5~\cite{Patzak:2001:OOFEM,OOFEM:WWW}. Spatial discretization is
performed using linear brick elements with eight integration
points. The arc-length method with elastic stiffness matrix is used to
determine the load-displacement curve related to the analyzed
experiment.

\section{PARAMETER IDENTIFICATION METHODOLOGY}
%%%%%%%%%%%%%%%%%%%%%%%%%%%%%%%%%%%%%%%%%%%%%%

The problem of model parameters identification starts by the design
and realization of a suitable experiment $E({\bf x}^\mathrm{E}) = {\bf y}^\mathrm{E}$,
which connects material properties ${\bf x}^\mathrm{E}$ with some observable
quantities ${\bf y}^\mathrm{E}$ under some given loading conditions. Then we
assume to have an appropriate numerical model $M({\bf x}^\mathrm{M}) = {\bf
  y}^\mathrm{M}$ approximating the experiment in an efficient and accurate way.
The goal is then to estimate the values of the model parameters ${\bf
  x}^\mathrm{M}$ minimizing the difference between the model outputs ${\bf
  y}^\mathrm{M}$ and observed data ${\bf y}^\mathrm{E}$.

As it was mentioned in introduction, many different identification
methodologies are available in literature. Nevertheless, only a
limited number of them were reliably verified. As the authors of
\cite{Babuska:2004:CMAME} introduced terms \bfit{verification} and
\bfit{validation} for numerical models of physical events, we will
follow here the definition of those terms in the field of parameters
identification presented in \cite{Kucerova:2007:PHD}:
\begin{itemize}
\item []
\bfit{Verification}: The process of determining whether the
identification method is able to~re-find the model parameters ${\bf x}^\mathrm{M}$
from the outputs ${\bf y}^\mathrm{ref}$ of the reference simulation done for any
choice of original inputs ${\bf x}^\mathrm{ref}$.
\item []
\bfit{Validation}: The process of determining whether the
identification method is able to find the model parameters ${\bf x}^\mathrm{M}$
corresponding to the experimental outputs ${\bf y}^\mathrm{E}$.
\end{itemize}

In this paper, we focus on an inverse mode of an identification strategy, which was already
verified for parameters identification on microplane model
\cite{Kucerova:2007:BAMM} and we would like to extend this work
by validation on experimental data. Therefore, we just briefly review
the main principles of the applied methodology and an interested
reader can find more information in \cite{Kucerova:2007:BAMM}.

The methodology assumes an existence of an inverse relationship
between model outputs ${\bf y}^\mathrm{M}$ and model inputs ${\bf
  x}^\mathrm{M}$, i.e. there is an inverse model $M_\mathrm{inv}$ associated
to the model $M$, which fulfils ${\bf x}^\mathrm{M} = M_\mathrm{inv}({\bf
  y}^\mathrm{M})$. Generally, this inverse model does not need to
exist. Nevertheless, we assume that the inverse model can be found
sufficiently precise on some closed subdomain of the parameter space.
Next, we will limit our attention to an approximation of the inverse
relationship, not its exact description. A quality of this
approximation is easy to measure since a pair ${\bf x}$, ${\bf y}$
obtained using inverse model $M_\mathrm{inv}(\cdot)$ should also fulfill the
forward model relation $M(\cdot)$.  Final usage of this methodology is
trivial because a desired value ${\bf x}^\mathrm{M}$ can be simply
obtained as ${\bf x}^\mathrm{M} = M_\mathrm{inv}({\bf y}^\mathrm{E})$.

The main advantage is clear. If an inverse relationship is
established, then the retrieval of desired inputs is a matter of
seconds even if executed repeatedly. This can be utilized for frequent
identification of one model. On the contrary, the main disadvantages
are an exhausting search for the inverse relationship, the existence
problems for the whole search domain and inability to solve the
problem of several global optima.

\begin{figure}[th]
\centering
\includegraphics[keepaspectratio,width=14cm]
{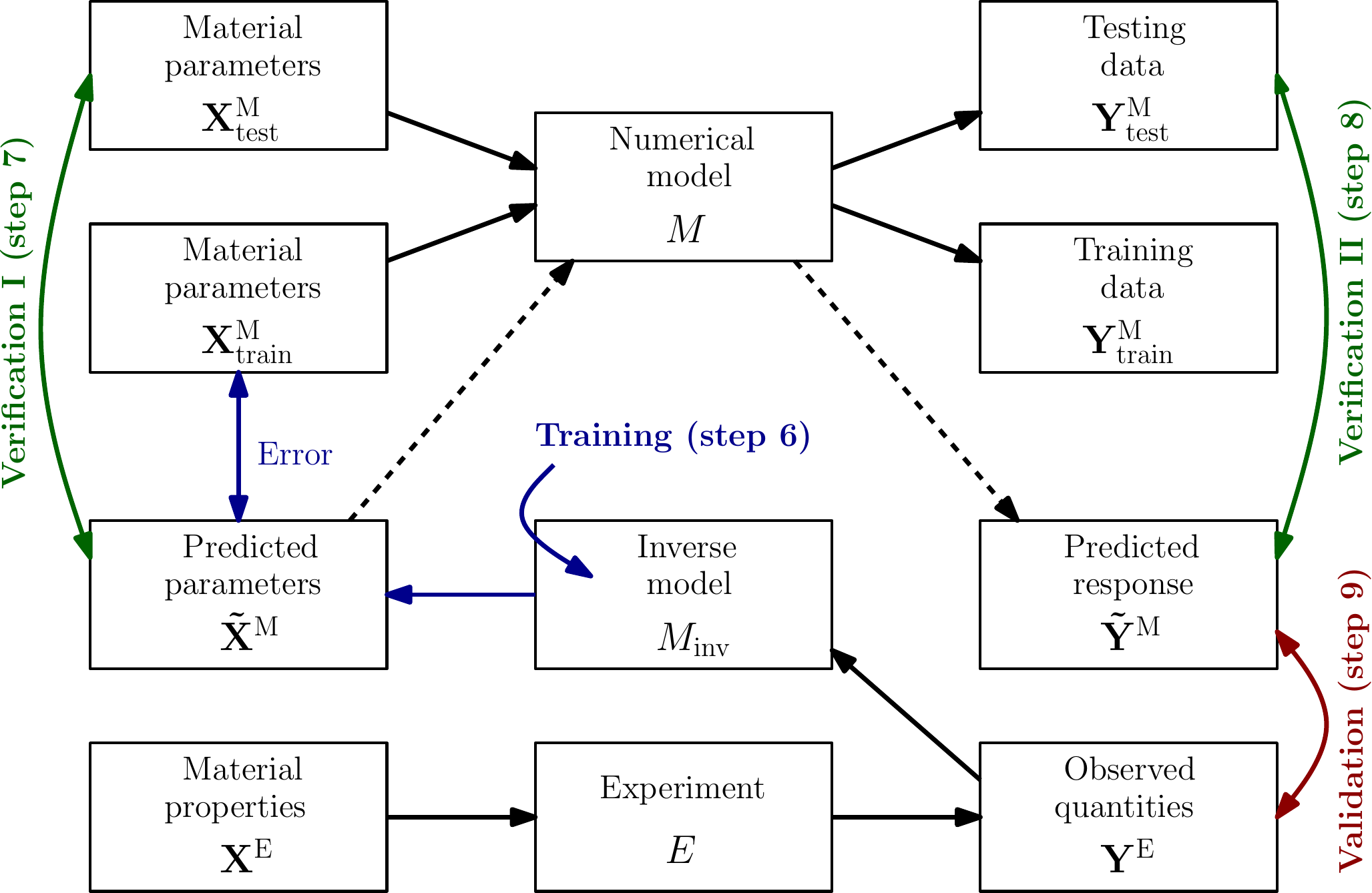}
\caption{Scheme of inverse analysis procedure}
\label{fig_inverse-schema}
\end{figure}

The presented identification strategy is based on an approximation of
inverse relation using an artificial layered neural network (ANN), which is known as a~very general and robust approximation tool. Individual steps of the
identification strategy involve:

\begin{description}

\item[Step~1]
\bfit{Setup of} an experimental \bfit{test} used for the identification
procedure and storing the observed quantities ${\bf y}^\mathrm{E}$.
\item[Step~2]
\bfit{Randomization} of input parameters. Input data are typically
assumed to be random variables uniformly distributed on a given
interval, which roughly lead to equally distributed precision of an ANN
over the parameter domain. Of course, any other distribution for
parameters is admissible to improve the ANN's accuracy e.g. in the vicinity of
particular parameter values. A~representative set of input vectors
${\bf X}^\mathrm{M}_\mathrm{train} = \left[{\bf x}_1^\mathrm{M}, {\bf x}_2^\mathrm{M},
\dots, {\bf x}_{n_\mathrm{train}}^\mathrm{M} \right]$ is carefully chosen for ANN training
following the design of experiments methodology~\cite{Montgomery:2005}. Another set of input
vectors ${\bf X}^\mathrm{M}_\mathrm{test} = \left[{\bf x}_1^\mathrm{M}, {\bf x}_2^\mathrm{M}, \dots,
{\bf x}_{n_\mathrm{test}}^\mathrm{M} \right]$ is randomly chosen for ANN testing. $n_\mathrm{train}$ and
$n_\mathrm{test}$ denote the number of training and testing samples, respectively.
\item[Step~3]
\bfit{Training and testing data} sets preparation.  The computational
model $M$ is applied to simulate the experiment $E$ for all training
and testing input vectors in order to obtain corresponding output
vectors ${\bf Y}^\mathrm{M}_\mathrm{train} = \left[{\bf y}_1^\mathrm{M}, {\bf y}_2^\mathrm{M}, \dots,
{\bf y}_{n_\mathrm{train}}^\mathrm{M} \right]$ and ${\bf Y}^\mathrm{M}_\mathrm{test} = \left[{\bf y}_1^\mathrm{M},
{\bf y}_2^\mathrm{M}, \dots, {\bf y}_{n_\mathrm{test}}^\mathrm{M} \right]$, respectively, where ${\bf
y}_i^\mathrm{M} = M({\bf x}^\mathrm{M}_i)$.
\item[Step~4]
Global \bfit{sensitivity analysis} using the training
simulations. This provides us with \bfit{relevant model parameters}
which can be reliably identified from the computational
simulation. Usually this step is performed by calculation the
correlation between inputs ${\bf X}^\mathrm{M}_\mathrm{train}$ and outputs ${\bf
Y}^\mathrm{M}_\mathrm{train}$.
\item[Step~5]
Definition of \bfit{the topology} of an ANN used for the identification
procedure.
\item[Step~6]
\bfit{Training} of the ANN, i.e. developing $M_\mathrm{inv}$. Usually
an optimization algorithm is needed to appropriately setup values of
synaptic weights ${\bf w}$ of the ANN by minimizing the error function
$E({\bf w}) = \| (M_\mathrm{inv}({\bf Y}^\mathrm{M}_\mathrm{train},{\bf w}) - {\bf X}^\mathrm{M}_\mathrm{train} \|$.
\item[Step~7]
\bfit{Verification I} of the ANN with respect to the computational
model. This step is usually performed by comparing the ANN's
prediction of model input parameters ${\bf {\tilde X}}^\mathrm{M}_\mathrm{test} =
M_\mathrm{inv}({\bf Y}^\mathrm{M}_\mathrm{test}, {\bf w})$, with the original one ${\bf X}^\mathrm{M}_\mathrm{test}$
for unseen testing (or reference) data.
\item[Step~8]
\bfit{Verification II} of the ANN with respect to the computational
model. In this step, a~computational model should be evaluated for
predicted values ${\bf {\tilde X}}^\mathrm{M}_\mathrm{test}$ in order to obtain
corresponding model outputs ${\bf {\tilde Y}}^\mathrm{M}_\mathrm{test}$. Then the
outputs ${\bf {\tilde Y}}^\mathrm{M}_\mathrm{test}$ are compared with the
original ones ${\bf Y}^\mathrm{M}_\mathrm{test}$. This step is not necessary, but is
utmost recommended.
\item[Step~9]
\bfit{Validation} of the ANN with respect to the experiment. The trained
ANN is evaluated for experimental data ${\bf y}^\mathrm{E}$ in order to obtain
corresponding input values ${\bf {\tilde x}}^\mathrm{E} = M_\mathrm{inv}({\bf y}^\mathrm{E})$
of the computation model $M$. The model $M$ is then evaluated for
obtained inputs ${\bf {\tilde x}}^\mathrm{E}$ and results ${\bf {\tilde y}}^\mathrm{E} =
M({\bf {\tilde x}}^\mathrm{E})$ are compared with original measured data ${\bf
y}^\mathrm{E}$.
\end{description}
For clarity, the scheme of such identification procedure is displayed
in Figure \ref{fig_inverse-schema}.

\subsection{NEURAL NETWORKS ARCHITECTURE}
%%%%%%%%%%%%%%%%%%%%%%%%%%%%%%%%%%%%%%%%%

Multilayer neural networks became very popular as an universal
approximation tool~\cite{Adeli:2001:NNCE}. The ANN is composed of a set of simple operational units -
neurons. Each neuron is represented by an activation function, which
defines the relation between an input and output value of the neuron. In
the multi-layered neural network, neurons are ordered into layers,
where outputs of neurons from one layer multiplied by
appropriate synaptic weights serve as input values for a neuron in the
next layer, see e.g.~\cite{Bishop:1995} for more details.

When looking for a suitable neural network, two non-trivial tasks must be solved~\cite{Yao:1999:EANN}: (a) a~choice of an appropriate architecture and (b) a~search for optimal values of synaptic weights, i.e. so-called training of a neural network. In our applications, we follow the premise that only three layers with enough hidden neurons are sufficient to describe any relation~\cite{Cybenko:1989}, therefore only three layers are used in our computations. Another premise used is that it is easier to train several networks with one output value than one complex network with several outputs~\cite{Kordik:2009:SCI}. Hence, one neural network is trained for each model parameter and the inverse model is then composed by a set of simple ANNs.

Another trick used in our computations is a \bfit{cascade neural network}. Here, a~sequential way of identification of the individual inputs $x_i$ is utilized to increase the prediction abilities of the ANN-based methodology by reducing the complexity of the approximated relationship. The predictions of some parameters identified in the preceding steps are assumed as known and serve as inputs during the development of inverse models in following steps. Several applications of this methodology to parameters identification are presented e.g. in \cite{Waszczyszyn:2005}.

Finally, a number of input neurons that corresponds to the number of relevant observable quantities should be selected. Since in many engineering problems, measurements are represented by single or several curves usually discretized into discrete points, a number of observed quantities is often very large. Therefore, a~global sensitivity analysis is applied to choose only quantities important for a~particular model parameter to be identified~\cite{Novak:2006:EAAN}. A number of neurons in hidden layer is determined by consecutively increasing their number taking the over-training and under-training issues into account. Particularly, a~goal is to find a~trade-off between errors in prediction for training and testing data, respectively.

\subsection{NEURAL NETWORK TRAINING}
%%%%%%%%%%%%%%%%%%%%%%%%%%%%%%%%%%%%

Once the architecture of the ANN is chosen, the training process can
start. The most famous training algorithm still remains a relatively simple
backpropagation, especially because of its simple interpretation and
implementation. As the name suggests, the error in prediction of the
output layer is propagated back to previous layers and according to
an error's value, synaptic weights are updated. From the optimization
point of view, it is a gradient-based algorithm, which suffers from premature
convergence.

Therefore, GRADE evolutionary algorithm extended by a niching strategy CERAF is used here as a more robust optimization algorithm. This algorithm was derived from its predecessor SADE algorithm, which provided better results in ANN training against the back-propagation as was presented in \cite{Drchal:2002:ECT}. GRADE is a population-based real-coded algorithm consisting of three genetic operators, which are applied to a set of solutions. Particularly, it involves mutation, differential cross-over and inverse tournament selection. CERAF is a multistart strategy enhanced with memory that increases the ability of a~population-based algorithm to escape from local extremes. An interested reader can find more details about both methods in \cite{Kucerova:2007:PHD} or \cite{GRADE:WWW, CERAF:WWW}, where C++ and MATLAB implementations are available. For information on other possible training algorithms see e.g.~\cite{Kordik:2009:SCI} or a review~\cite{Yao:1999:EANN}.

\section{EXPERIMENTAL VALIDATION}
%%%%%%%%%%%%%%%%%%%%%%%%%%%%%%%%%

Following the heuristic calibration procedure
suggested in~\cite{Caner:2000:MM4II}, we examine three specific
experimental tests: (i)~uniaxial compression, (ii)~hydrostatic test
and (iii)~triaxial test. Advantage of these tests is their simplicity
and availability in most experimental facilities. Moreover, authors
in~\cite{Caner:2000:MM4II} claim that these experiments are sufficient
to determine all parameters of the microplane model M4. The results
presented in this section can be understood as a~verification of this
claim.

In our previous contribution~\cite{Kucerova:2007:BAMM}, we have shown
that the proposed methodology is able to identify all parameters from
computer-simulated curves. To demonstrate the applicability of the
proposed procedure, a~real experiment should be examined. Since the
measurements from different loading tests are obtained for different
concretes, this section does not represent a validation of proposed
identification strategy in general, but only the validation of
application of particular inverse models.

Each neural network is obtained by the procedure described previously. More precisely, a fully connected three-layer perceptron with bias neurons is applied to map discrete values from stress-strain diagrams to microplane model parameters. Log-sigmoid functions are considered as activation functions in all neurons. In all cases studied in this work, 60 training samples are generated by Latin Hypercube Sampling method and optimized by Simulated Annealing in order to minimize the correlation among all samples~\cite{Novak:2006:EAAN}. Those two methods are implemented in FREET software \cite{Novak:2003:FREET} that has been used. Next, 10 testing samples are obtained for random parameters from given bounds. Relevant values selected from stress-strain diagram as ANN's inputs are chosen by hand with respect to the results of global sensitivity analysis. Here, the Pearson product-moment correlation coefficient~\cite{Holmes:2001:Corr} is evaluated between the discrete values of stresses (or strains) and corresponding values of microplane model parameters. For the ANN training the GRADE algorithm supported by CERAF strategy is used and
calculation is stopped after 1,000,000 iterations of the
algorithm. Finally, all three following compression tests are performed on the cylinders with a~radius equal to 75~mm and the height of 300~mm.

\subsection{UNIAXIAL COMPRESSION TEST}\label{sec:uni}
%%%%%%%%%%%%%%%%%%%%%%%%%%%%%%%%%%%%%%

\begin{figure}[h!]
\centering
\begin{tabular}{cc}
\includegraphics[width=0.45\textwidth,keepaspectratio]{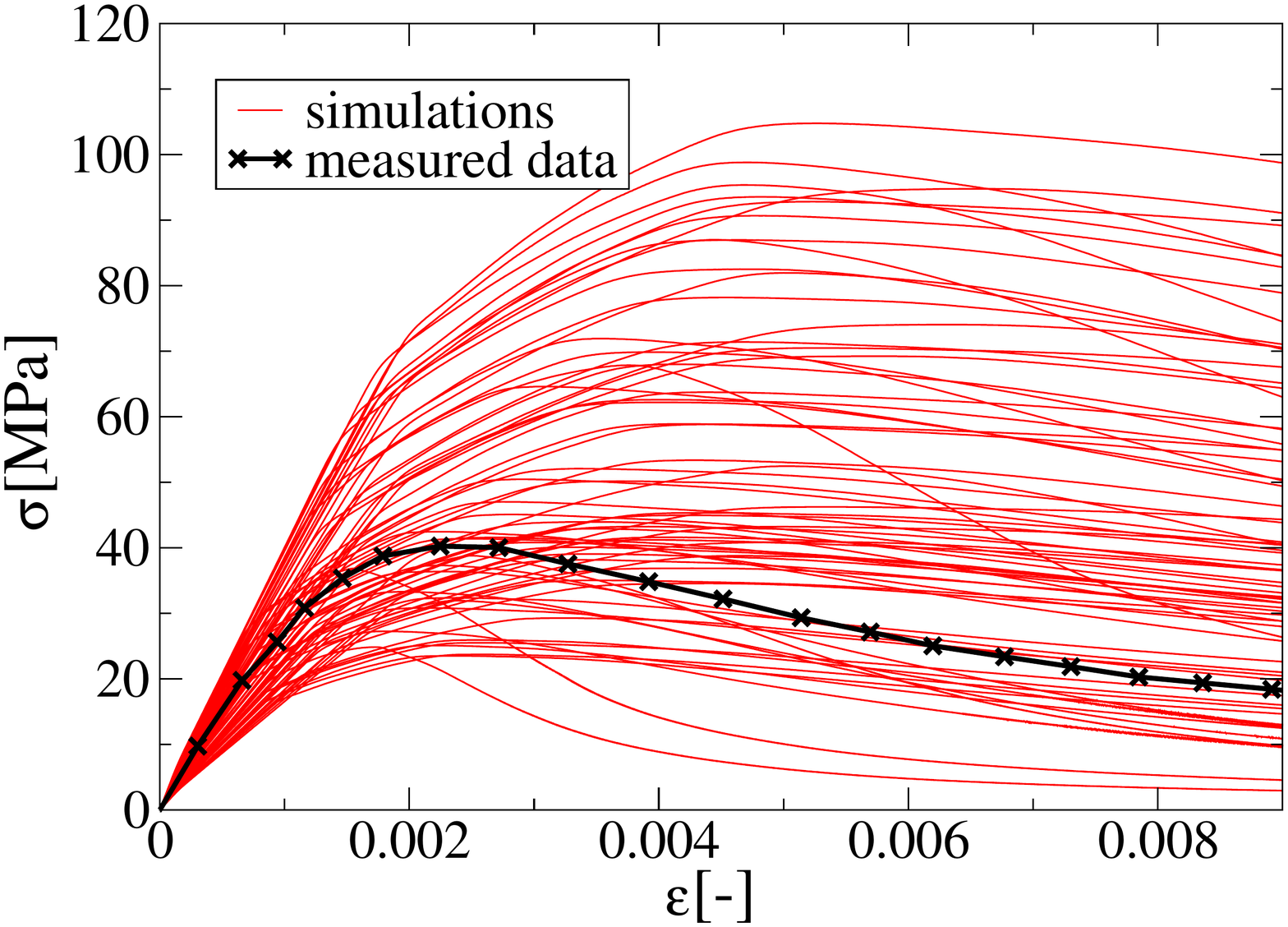} &
\includegraphics[width=0.45\textwidth,keepaspectratio]{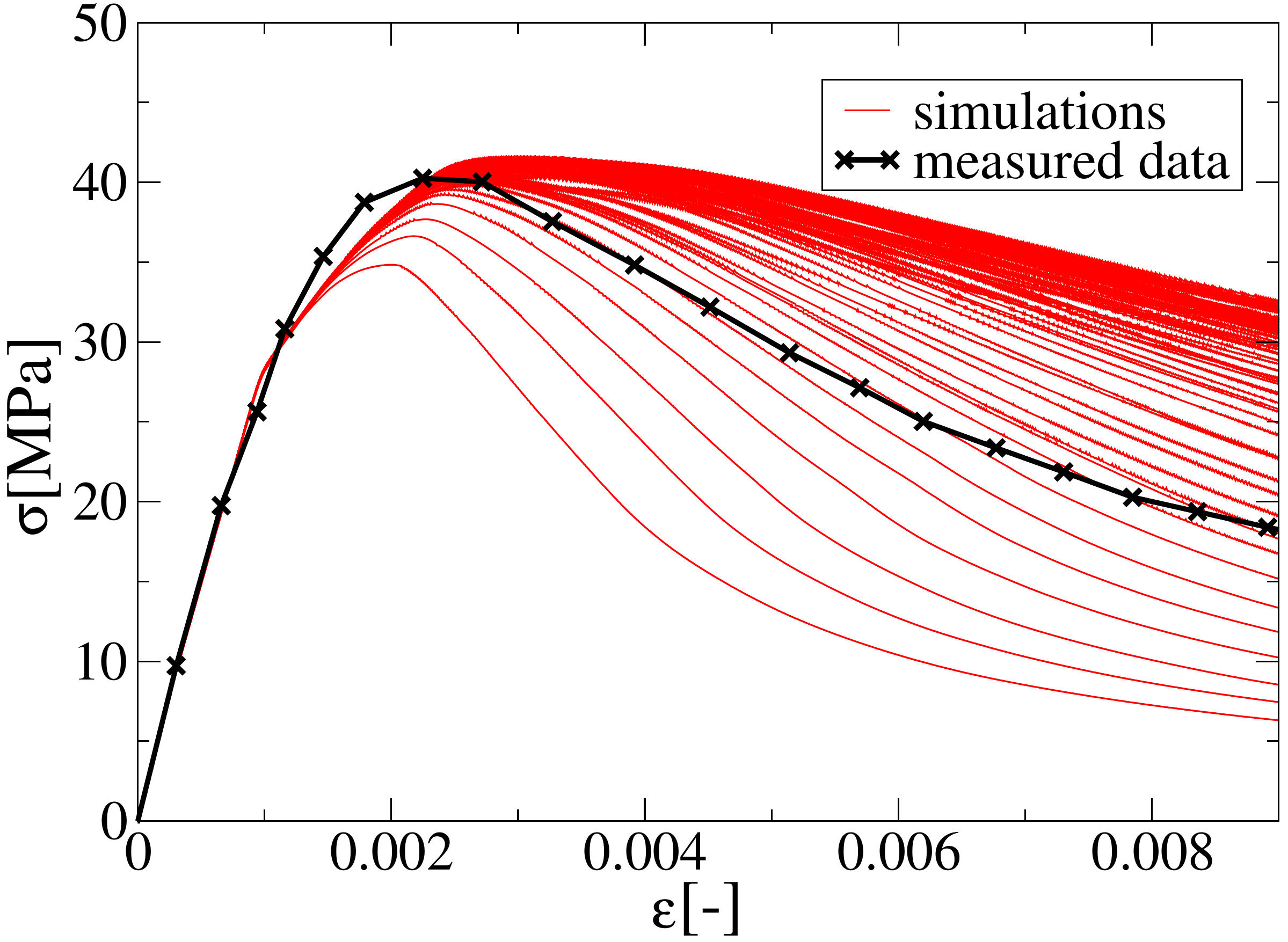} \\
(a) & (b)
\end{tabular}
\caption{Bundle of curves of uniaxial compression test simulation with
  experimental data for varying (a) all parameters and (b) only the
  constant $c_{20}$.}
\label{fig_uni_bundle_all}
\end{figure}

\begin{figure}[!ht]
\centering
\begin{tabular}{cc}
\includegraphics[keepaspectratio,width=0.495\textwidth]{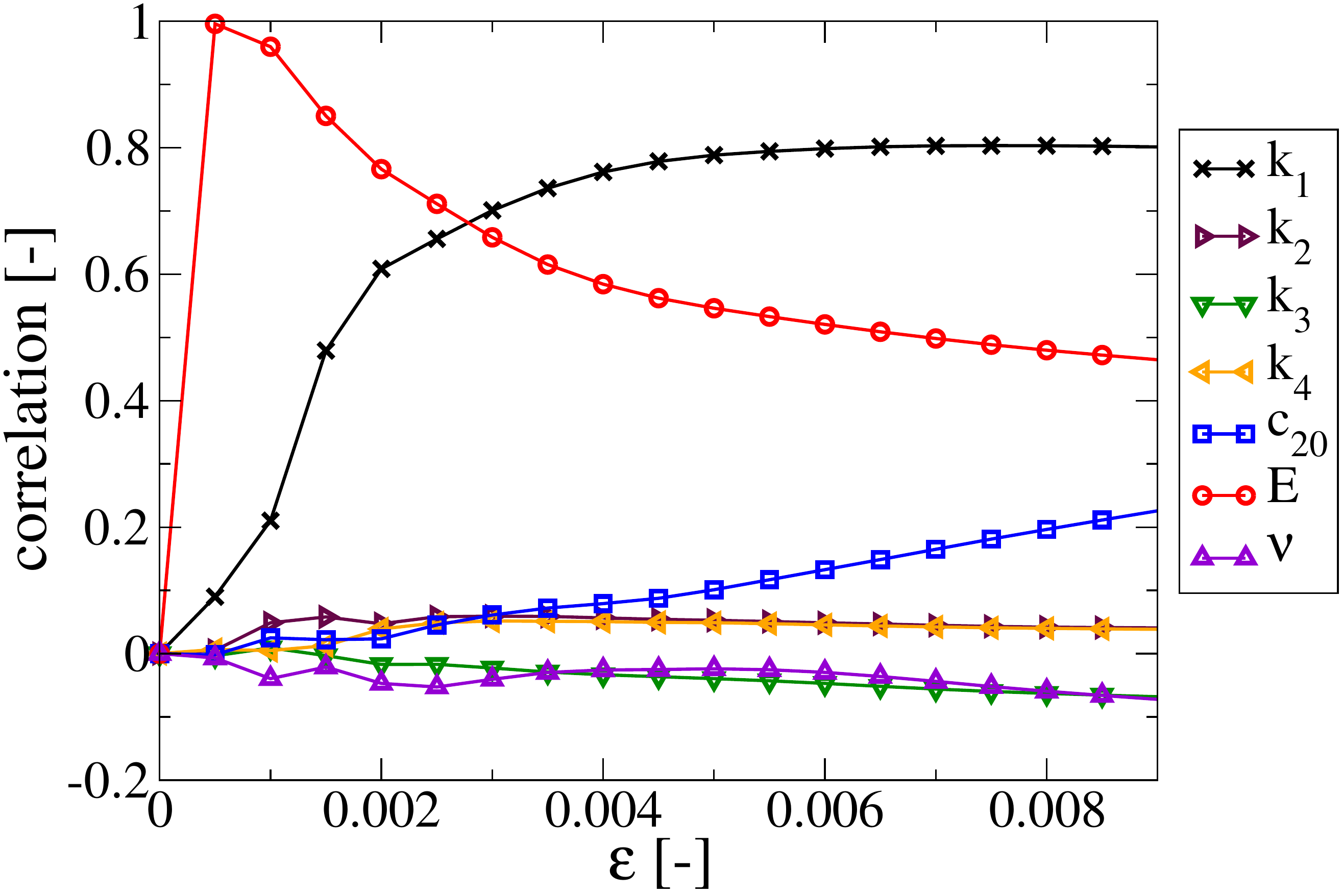} &
\includegraphics[keepaspectratio,width=0.385\textwidth]{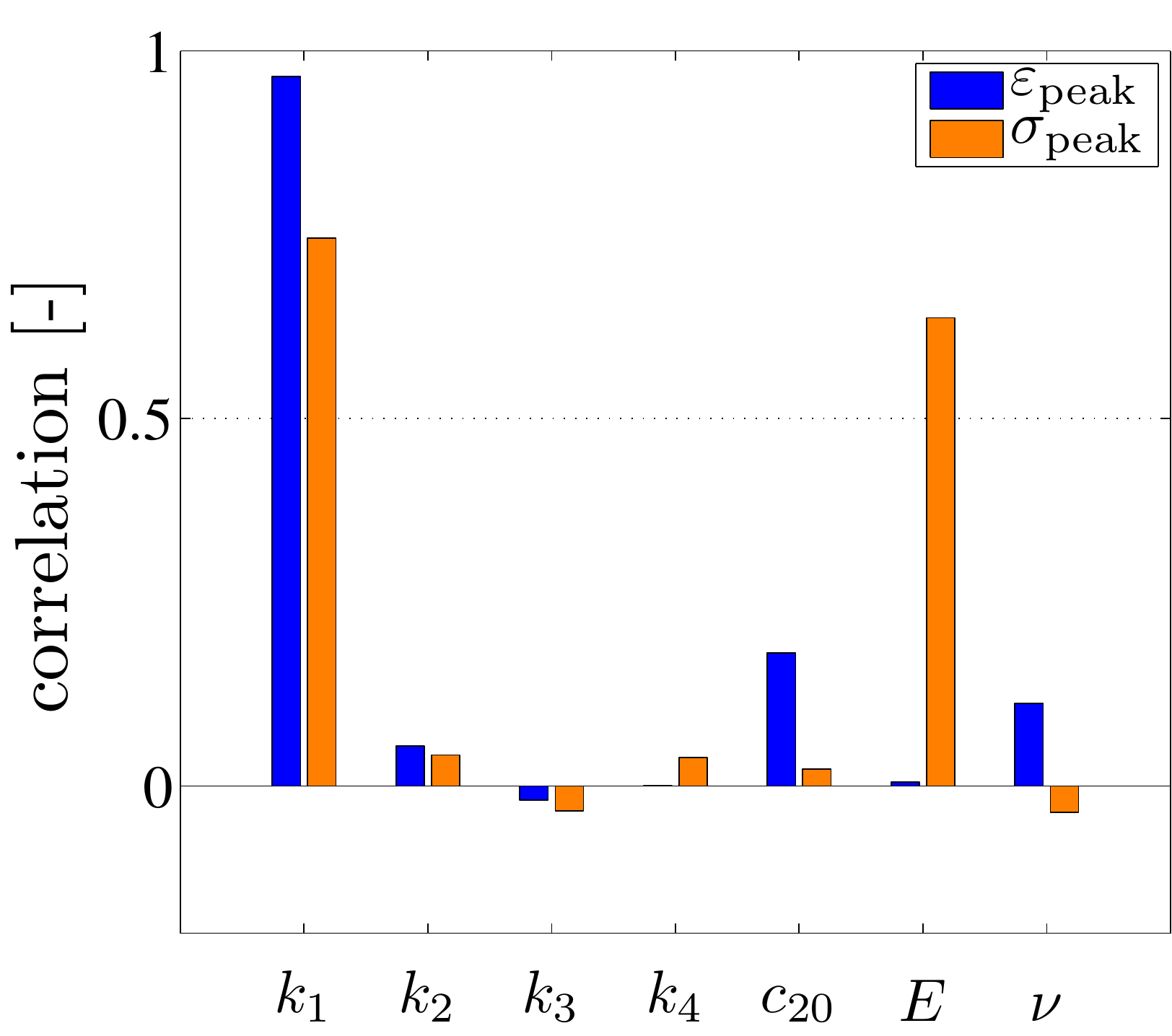} \\
(a) & (b)
\end{tabular}
\caption{Sensitivity analysis for uniaxial compression test: (a)
  evolution during the loading process and (b) peak coordinates
  [$\epsilon_\mathrm{peak}$,$\sigma_\mathrm{peak}$] of stress-strain
  curves.}
\label{fig:uniaxial_compression_sensitivity}
\end{figure}

Experimental data are taken from \cite[Figure 1a]{Caner:2000:MM4II}.
Note that the original source comes from \cite{vanMier:1984}. A suite
of 70 training and testing simulations and the resulting bundle of
curves together with the experimental data is shown in
Figure~\ref{fig_uni_bundle_all}a.
Only for illustrative purposes, the evolution of global sensitivity
during the loading process (horizontal axis) is depicted in
Figure~\ref{fig:uniaxial_compression_sensitivity}a and is omitted for
other two tests. The results indicate that the most sensitive
parameters are Young's modulus $E$, the coefficient $k_1$ and, for the
later stages of loading, the coefficient $c_{20}$. Therefore, one can
expect that only these parameters can be reliably identified from this
test. Poisson's ratio $\nu$ can be also identified from this type of
experiment, but only if lateral deformation is measured.

Moreover, the impact of individual parameters on a position of a peak
of stress-strain curves is computed. The results of a~sensitivity
analysis using Pearson's product moment correlation coefficient of
peak coordinates [$\epsilon_\mathrm{peak}$,$\sigma_\mathrm{peak}$] are plotted in
Fig~\ref{fig:uniaxial_compression_sensitivity}b.  Results indicate particularly strong
influence of the $k_1$ parameter, which promise its reliable
determination.

Based on the results of sensitivity analysis, the neural network training can be performed using a~nested strategy of cascade neural networks. First, Young's modulus $E$ with sensitivity $\approx 1$ in the initial stage is easily identified. To this end, the following ANN's inputs are chosen: the values of stresses $\sigma_{[\hat{\epsilon}]}$, where $\hat{\epsilon}$ is a particular level of deformation with extremal Pearson's correlation coefficient, here the starting of the loading curve, particularly $\hat{\epsilon} = 0.0005; 0.001$ and $0.0015$, respectively. The hidden layer contains two neurons only; the output layer consists of one neuron corresponding to the predicted value of Young's modulus $E$, see also notation used in Table~\ref{tab_uni_ann_top}. Similar approach is used for $k_1$ parameter with the architecture described in Table~\ref{tab_uni_ann_top}. Note the usage of Young's modulus $E$ estimation from the previous ANN. Errors for both ANN's predictions are then listed in Table~\ref{tab_uni_ann_err}.

\begin{table}[ht!]
\centering
\begin{tabular}{ccc}
\hline
\textbf{Parameter} & \textbf{Topology} & \textbf{Inputs} \\
\hline
$E$ & $3+2+1$ & $\sigma_{[0.0005]}, \sigma_{[0.001]}, \sigma_{[0.0015]}$ \\
$k_1$ & $5+3+1$ & $\sigma_{[0.0025]}, \sigma_{[0.009]}, \epsilon_\mathrm{peak}, \sigma_\mathrm{peak}$, prediction of $E$ \\
\hline
\end{tabular}
\caption{Topology of neural networks for uniaxial compression test.}
\label{tab_uni_ann_top}
\end{table}

\begin{table}[ht!]
\centering
\begin{tabular}{c|cc|cc}
\hline
\textbf{Parameter} & \multicolumn{2}{c}{\textbf{Training data}} & \multicolumn{2}{c}{\textbf{Testing data}} \\
 & Maximal error & Average error & Maximal error & Average error \\
\hline
$E$ & 1.97 & 1.01 & 1.50 & 0.81 \\
$k_1$ & 2.68 & 1.57 & 2.63 & 1.50 \\
\hline
\end{tabular}
\caption{Errors in ANN's prediction relative to the size of admisible domain given for the parameters in [\%].}
\label{tab_uni_ann_err}
\end{table}

Next, the trained neural networks were applied on measured data and
the following values were obtained: $E = 36057$ MPa and $k_1 =
0.000196$, while the M4 authors suggest in \cite{Caner:2000:MM4II} $E
= 32173$ MPa and $k_1 = 0.000165$.

Since the lateral deformation is missing in measured data, Poison's
ratio cannot be identified. Its value is chosen according to
microplane authors' recommendations $\nu = 0.2$. According to
sensitivity analysis, other parameters should not be important.
Therefore their values are set to mean values of predefined intervals:
$k_2 = 550$, $k_3 = 10$ and $k_4 = 115$.\footnote{For the sake of
  completeness, we should note that the M4 authors suggest different
  values for the two parameters: $k_2 = 160$ and $k_4 = 150$}

\begin{figure}[ht!]
\vspace{1cm}
\centering
\includegraphics[keepaspectratio,width=0.8\textwidth]{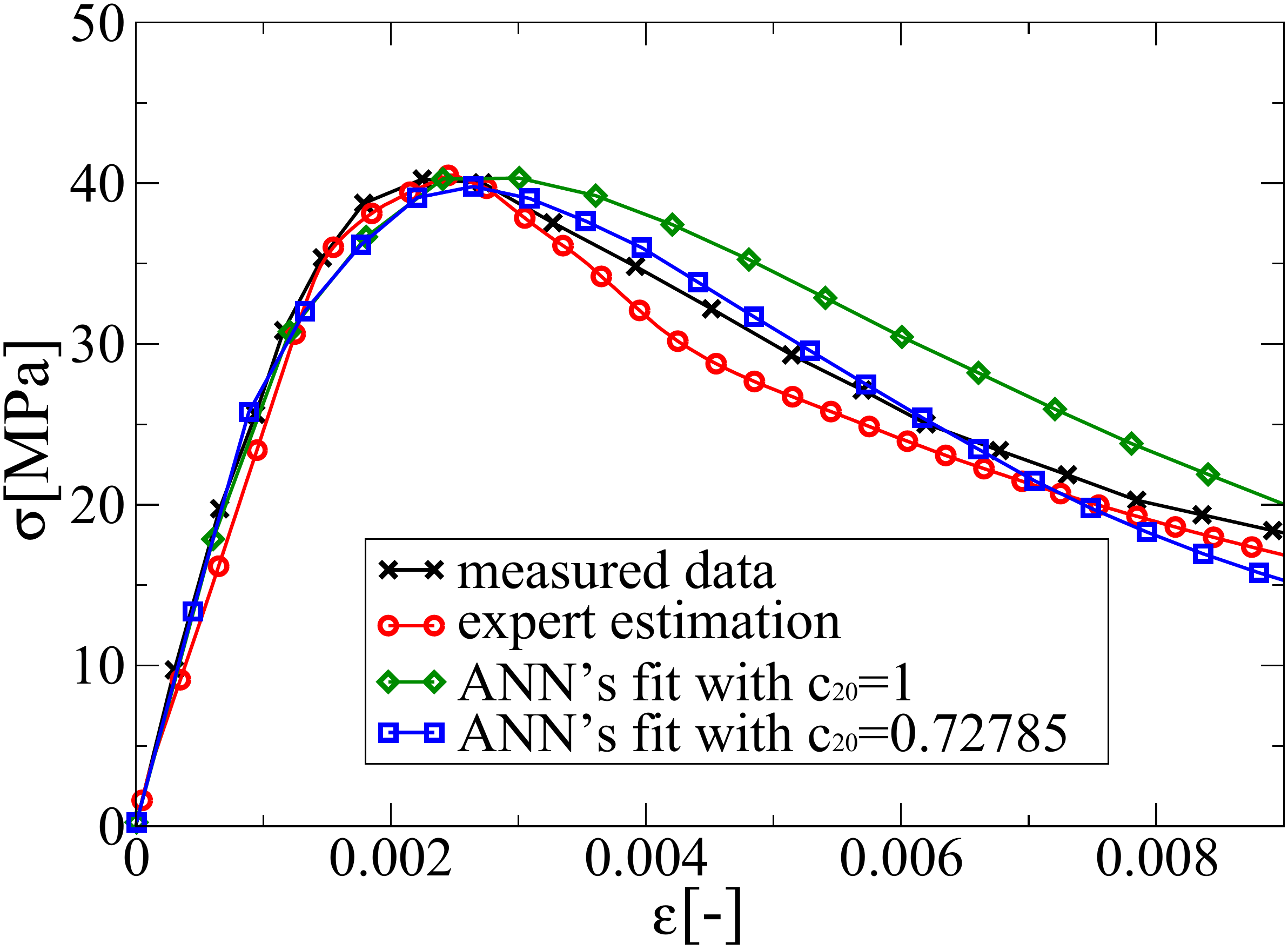}
\caption{Comparison of measured data and predicted simulations for uniaxial compression test.}
\label{fig_uni_c20_comp}
\end{figure}

Finally, \new{the} value of \new{the} constant $c_{20}$ recommended by
M4 authors is $c_{20} = 1.0$. In Figure~\ref{fig_uni_c20_comp},
measured data are compared to the simulation obtained for \new{the}
predicted parameters and to the simulation obtained by M4 authors
using \new{a} hand-fitting method.

Following our experience~\cite{Nemecek00:73353}, it is interesting to
predict also the value of \new{the} constant $c_{20}$. Therefore, new
70 simulations (60 for training and 10 for testing purposes,
respectively) were performed with fixed values of $E$, $\nu$ and $k_1$
parameters. The resulting bundle of curves can be compared to measured
data, see Figure~\ref{fig_uni_bundle_all}b. The same discretization
was respected. The topology and inputs for the neural network are
presented in Table~\ref{tab_uni_c20_ann_top} and errors in ANN's
predictions are listed in Table~\ref{tab_uni_c20_ann_err}.

\begin{table}[ht!]
\centering
\begin{tabular}{ccc}
\hline
\textbf{Parameter} & \textbf{Topology} & \textbf{Inputs} \\
\hline
$c_{20}$ & $4+2+1$ & $\sigma_{[0.003]}, \sigma_{[0.004]}, \sigma_{[0.006]}, \sigma_{[0.008]}$ \\
\hline
\end{tabular}
\caption{Topology of neural network for constant $c_{20}$ identification.}
\label{tab_uni_c20_ann_top}
\end{table}

\begin{table}[ht!]
\centering
\begin{tabular}{c|cc|cc}
\hline
\textbf{Parameter} & \multicolumn{2}{c}{\textbf{Training data}} & \multicolumn{2}{c}{\textbf{Testing data}} \\
 & Maximal error & Average error & Maximal error & Average error \\
\hline
$c_{20}$ & 11.18 & 6.02 & 11.27 & 6.15 \\
\hline
\end{tabular}
\caption{Errors in ANN's prediction of constant $c_{20}$ relative to the size of admisible domain given for the $c_{20}$ parameter in [\%].}
\label{tab_uni_c20_ann_err}
\end{table}

Neural networks were applied on measured data and following value was obtained:
$c_{20} = 0.72785$.

The final comparison of measured data and predicted simulations is shown in Figure~\ref{fig_uni_c20_comp}. To show some more objective comparison of presented simulations, the error between measured data and simulated curves can be summarized over discrete points corresponding to measured data, i.e.
\begin{equation}
E = \| \tenss{\stress} - \tilde{\tenss{\stress}} \| ,
\label{eq_val-uni-err}
\end{equation}
where $\tenss{\stress} = [\dots, \sigma_i, \dots]$ corresponds to stresses measured in discrete points $i$ and $\tilde{\tenss{\stress}} = [\dots, \tilde{\sigma}_i, \dots]$ are corresponding simulated values. Values of the error defined by Equation~\ref{eq_val-uni-err} are shown in Table~\ref{tab_val-uni-err}.

\begin{table}[!ht]
\centering
\begin{tabular}{lc}
\hline
\textbf{Prediction} & \textbf{Error} $E$ \\
\hline
expert estimation \cite{Caner:2000:MM4II} & 8.30 \\
ANN + $c_{20}=1$ & 12.88 \\
ANN + $c_{20}=0.72785$ & 6.97 \\
\hline
\end{tabular}
\caption{Comparison of errors of predicted simulations.}
\label{tab_val-uni-err}
\end{table}

\subsection{HYDROSTATIC COMPRESSION TEST}
%%%%%%%%%%%%%%%%%%%%%%%%%%%%%%%%%%%%%%%%%

%Note that the maximal
%value of a hydrostatic pressure for all these tests is 427.5 MPa.

The next independent test used for the identification problem is the
hydrostatic compression test, where a concrete cylinder is subjected
to an increasing uniform pressure. Experimental data come again
from~\cite{Caner:2000:MM4II}. These data were obtained by authors
Green and Swanson~\cite{Green:1973:SCRC}. The stress-strain diagram
represents the relation of hydrostatic pressure $\sigma$ and axial
deformation $\varepsilon$. The results from simulation performed for
microplane model parameters obtained by expert estimation are there
also available in comparison with measured data.

\begin{figure}[!ht]
\centering
\begin{tabular}{cc}
\includegraphics[width=0.45\textwidth,keepaspectratio]{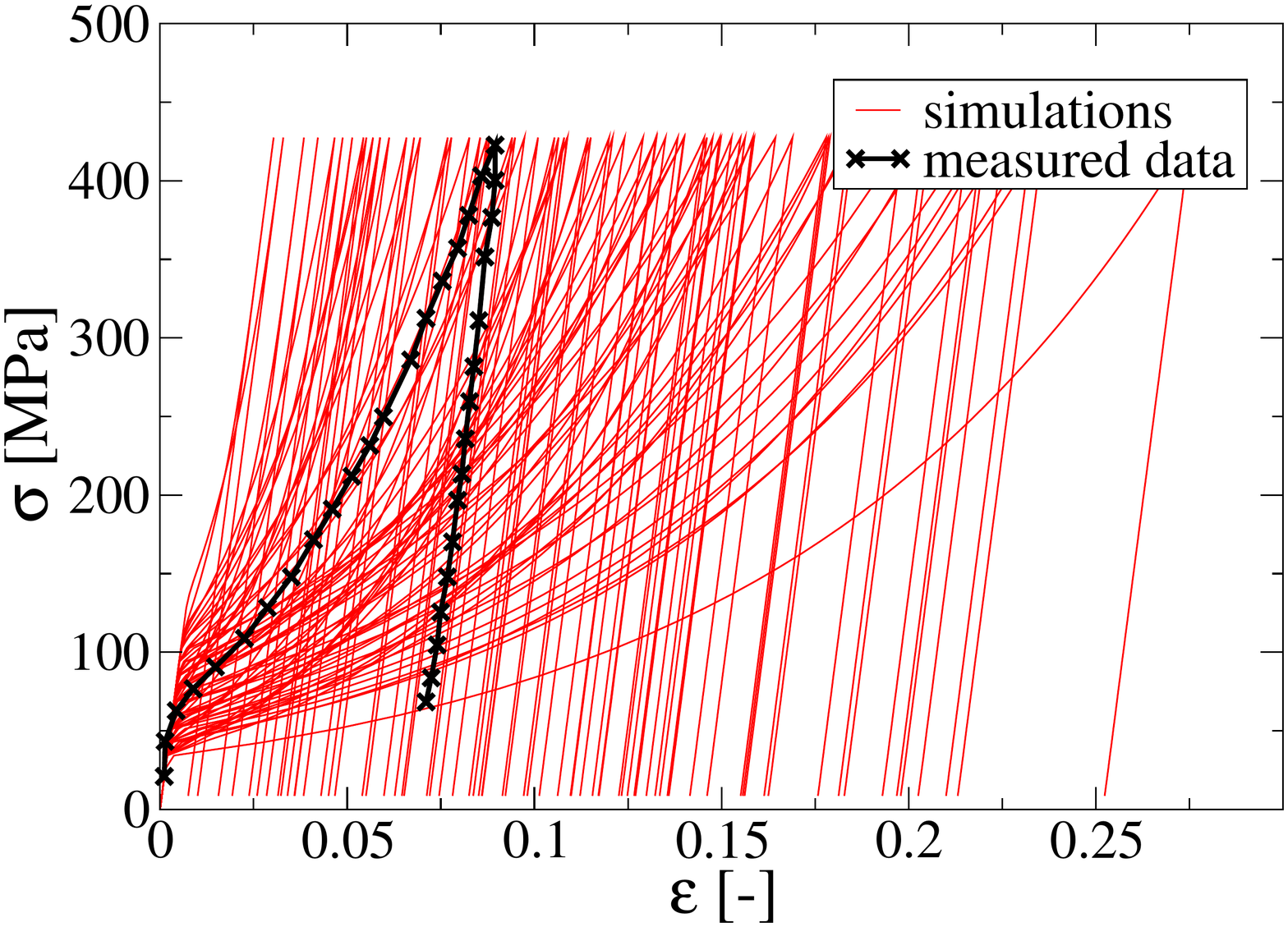} &
\includegraphics[width=0.40\textwidth,keepaspectratio]{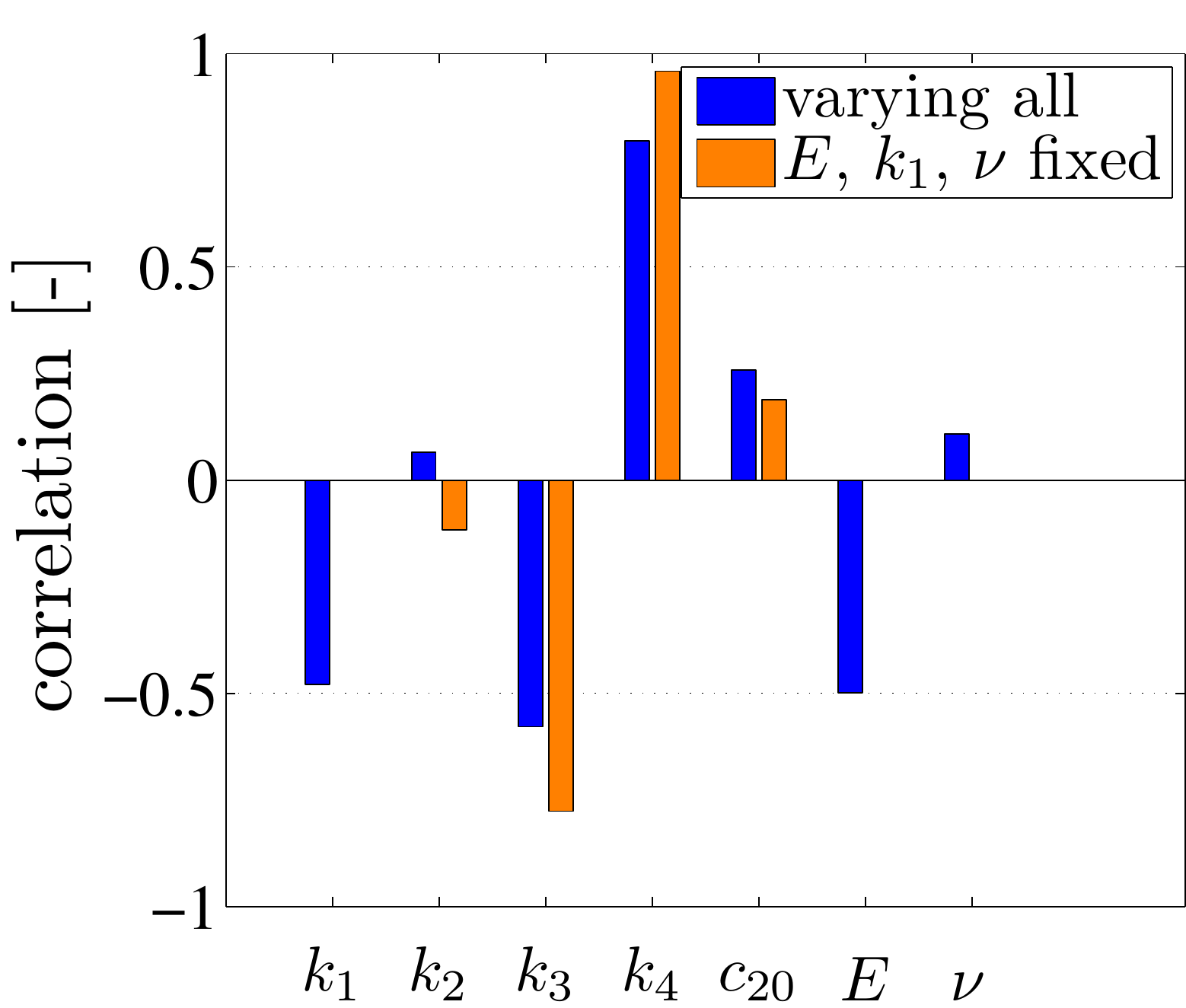} \\
(a) & (b)
\end{tabular}
\caption{Hydrostatic compression test: (a) comparison of measured data and results of 70 simulations; (b) maximal sensitivities.}
\label{fig_val-hyd-bundle}
\end{figure}

\new{We suppose that values of Young's modulus, Poisson's ratio and
  $k_1$ parameter can be reliably obtained from uniaxial compression
  test. If not, they cannot be easily estimated from hydrostatic
  compression as their sensitivity to this experiment is relatively
  low, see Figure ~\ref{fig_val-hyd-bundle}b, where maximal achieved
  sensitivities are plotted for the case of varying all the microplane
  parameters and the case with their fixed (known) values.
  Nevertheless, their low influence on the material behavior is still
  high enough to deteriorate the sensitivity of $k_3$ and $k_4$
  parameters which may be then harder to identify.}

Since the values of Young's modulus, Poisson's ratio and $k_1$
parameter are nevertheless not available for the concrete observed
here, their values are taken directly from~\cite{Caner:2000:MM4II}.
Then, the goal is to identify values of parameters $k_3$ and $k_4$;
parameter $k_2$ attains almost zero sensitivity here. The bundle of
resulting stress-strain diagrams for training and testing sets can be
compared with measured data in Figure~\ref{fig_val-hyd-bundle}a.

\begin{table}[!ht]
\centering
\begin{tabular}{c|c|c}
\hline
\bf Parameter & \bf ANN's layout & \bf Input values\\
\hline
$k_4$    & 3 + 2 + 1  & $k_3$, $\epsilon_\mathrm{peak}$, $\epsilon_{[85.5],\mathrm{unload}}$   \\
\hline
\end{tabular}
\caption{Neural network architecture for $k_{4}$ parameter identification; $\epsilon_{[85.5],u}$ is the deformation corresponding to hydrostatic stress $\hat{\stress} = 85.5$ MPa in the unloading phase.}
\label{tab_hyd-ann}
\end{table}

During the verification process, we have not been able to properly
identify the $k_3$ and $k_4$ parameters. Since all other parameters
have very small or zero correlation, we suppose that the nonlinearity
in the parameter $k_3$ identification is caused by $k_4$ parameter and
vice-versa. In other words, these difficulties can be caused by some
level of correlation between the parameters $k_3$ and $k_4$.
Therefore, to eliminate unknown correlation between \new{the}
parameters $k_3$ and $k_4$, their values are used also as inputs into
ANNs.

All the inputs and architecture of the ANN for predicting $k_4$
parameter are given in Table~\ref{tab_hyd-ann}. The inputs to an ANN
for prediction of $k_3$ parameter were chosen in two different ways in
order to show an impact of \new{their particular selection to
  prediction} of the post-elastic loading part of the stress-strain
diagram.  Two neural networks were trained with the deformation value
(i) in the middle of loading phase with stress $\hat{\stress} = 214$
MPa and (ii) in its thirds: $\hat{\stress} = 137$ MPa and
$\hat{\stress} = 308$ MPa leading to topologies given in
Table~\ref{tab_val-hyd-k3-ann}.  Note that the second ANN puts more
emphasis on the post-elastic part of the diagram than the first one.
\begin{table}
\centering
\begin{tabular}{ccc}
\hline
 & \textbf{Topology} & \textbf{Inputs} \\
\hline
ANN$_1$ & $4 + 2 + 1$ & $k_4$, $\epsilon_\mathrm{peak}$, $\epsilon_\mathrm{yield}$, $\epsilon_{[214],\mathrm{load}}$ \\
ANN$_2$ & $5 + 2 + 1$ & $k_4$, $\epsilon_\mathrm{peak}$, $\epsilon_\mathrm{yield}$, $\epsilon_{[137],\mathrm{load}}$, $\epsilon_{[308],\mathrm{load}}$ \\
\hline
\end{tabular}
\caption{Description of two neural networks trained to predict $k_3$ parameter}
\label{tab_val-hyd-k3-ann}
\end{table}

\begin{table}
\centering
\begin{tabular}{c|cc|cc}
\hline
\textbf{Parameter} & \multicolumn{2}{c|}{\textbf{Training data}} & \multicolumn{2}{c}{\textbf{Testing data}} \\
 & Average error & Maximal error & Average error & Maximal error \\
\hline
$k_3$ & 1.40 & 2.59 & 1.71 & 3.07 \\
$k_4$ & 1.51 & 2.52 & 1.21 & 2.13 \\
\hline
\end{tabular}
\caption{Error in ANN's predictions relative to the size of admisible domain given for the parameters in [\%].}
\label{tab_val-hyd-errk3k4}
\end{table}
Errors in predictions for training and testing data are listed in
Table~\ref{tab_val-hyd-errk3k4}. For the sake of brevity only
\new{the} errors corresponding to the first ANN trained for $k_3$
parameter are presented. The errors of the second network were
negligibly higher.

\begin{figure}[!ht]
\centering
\includegraphics[width=0.8\textwidth,keepaspectratio]{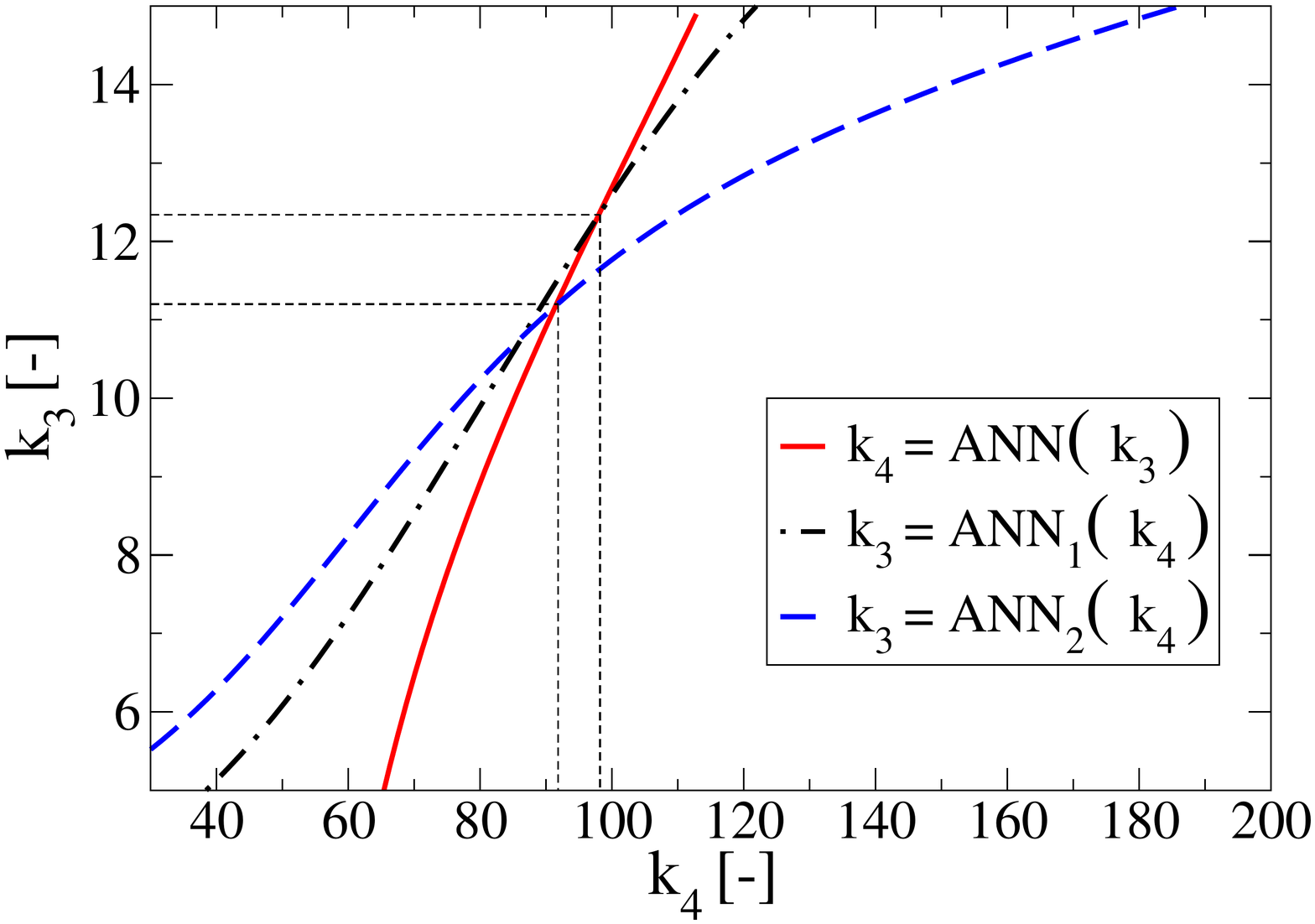}
\caption{Relations of $k_3$ and $k_4$ parameters for measured data: red continuous curve represents ANN trained to predict $k_4$ parameter, black dash-and-dot curve corresponds to ANN trained to predict $k_3$ parameter with four inputs (ANN$_1$) and blue dash curve to five inputs (ANN$_2$).}
\label{fig_val-hyd-k3k4}
\end{figure}

Two neural networks obtained to predict \new{the} $k_3$ and $k_4$
parameters with inputs taken from measured data represent system of
two non-linear equations, which can be solved numerically. Solutions
of particular equations are shown in Figure~\ref{fig_val-hyd-k3k4}.
The intersection of presented curves (i.e. solution of both equations)
defines the predicted values of $k_3$ and $k_4$ parameters
corresponding to measured data, in this case $k_3 = 12.34$, $k_4 =
98.19$ and $k_3 = 11.20$, $k_4 = 91.85$ for first and second neural
networks, respectively.

\begin{figure}[!ht]
\centering
\includegraphics[width=0.8\textwidth,keepaspectratio]{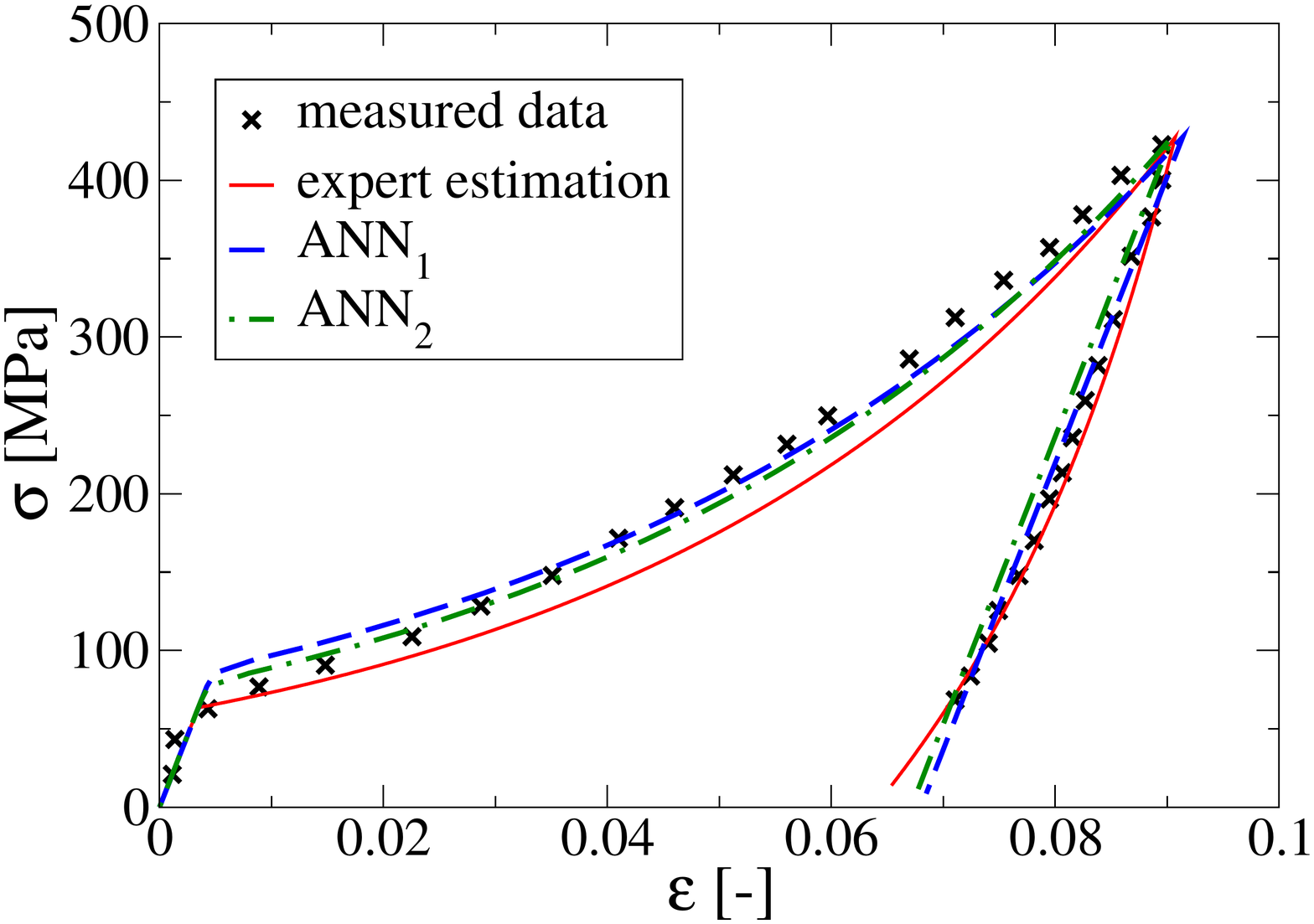}
\caption{Comparison of measured data and simulated diagrams of hydrostatic compression test for predicted parameters.}
\label{fig_val-hyd-comp}
\end{figure}

In both cases, the predicted values differ from values given by
authors in \cite{Caner:2000:MM4II}, where $k_3 = 9$ and $k_4 = 82$.
The comparison of measured data with simulated diagrams obtained for
parameters given in \cite{Caner:2000:MM4II} and for two couples of
parameters predicted by neural networks is depicted in Figure
\ref{fig_val-hyd-comp}.

It is not so easy to judge which predicted simulation really better
corresponds to measured data. A~simulation presented in
\cite{Caner:2000:MM4II} relatively well fits the measured data at the
end of elasticity and at the end of loading. Nevertheless, there is
a~significant error in the middle of the loading part of the diagram.
One can notice that the accuracy of ANN's prediction correspond to the
choice of ANN's inputs. The~simulation for the ANN$_1$ prediction fits
relatively well the measured data in the middle of the loading
diagram, while the simulation for the ANN$_2$ prediction fits the data
better near to the end of elasticity and to the end of the loading.
To show more objective comparison of presented simulations, the error
between measured data and simulated curves can be evaluated similarly
to Eq.~(\ref{eq_val-uni-err}) just in terms of deformation instead of
stresses.  The values of the resulting error are presented in
Table~\ref{tab_val-hyd-err}.
\begin{table}[!ht]
\centering
\begin{tabular}{lc}
\hline
\textbf{Prediction} & \textbf{Error} $E$ \\
\hline
expert estimation \cite{Caner:2000:MM4II} & 0.0257 \\
ANN$_1$ & 0.0163 \\
ANN$_2$ & 0.0161 \\
\hline
\end{tabular}
\caption{Comparison of errors of predicted simulations.}
\label{tab_val-hyd-err}
\end{table}

From the comparison presented in Table~\ref{tab_val-hyd-err} it is
clearly visible that simulations performed for parameters predicted by
both neural networks fit the measured data better than the simulation
done for parameters obtained by expert estimation presented
in~\cite{Caner:2000:MM4II}. Moreover, it is also visible that
simulations predicted by neural networks are somehow handicapped.
\new{ Particular implementation of the microplane model M4 in the
  OOFEM software is slightly simplified and thus it does not properly
  describe material behavior during the hydrostatic unloading. While
  the full microplane model accounts for nonlinear response within the
  unloading phase, the version in OOFEM software assumes here only
  linear behavior. This model imperfection of course induce higher
  error in the experimental data fits obtained for ANNs' predictions
  and one can expect that for the full microplane model simulations
  ANNs will achieve even better results. }
\subsection{TRIAXIAL COMPRESSION TEST}\label{sec_m4-val-tri}
%%%%%%%%%%%%%%%%%%%%%%%%%%%%%%%%%%%%%%%%%%%%%%%%%%%%%%%%%%%%

The last experiment, used for the purpose of parameter identification,
is a~triaxial compression test. To this end, a~specimen is subjected
to the hydrostatic pressure~$\stress_H$. After the peak value
of~$\stress_H$ is reached, the axial stress is proportionally
increased. The ``excess'' axial strain~$\strain = \strain_T -
\strain_H$, where $\strain_T$ and $\strain_H$ denote the total and
hydrostatic axial strain, is measured as a~function of the overall
stress~$\stress$. Similarly to the hydrostatic compression test, also
for triaxial compression test we use measured data from
\cite{Caner:2000:MM4II}. These data were obtained by
Balmer~\cite{Balmer:1949}. Again, the presented measured data are
accompanied by the simulation performed by the authors of
\cite{Caner:2000:MM4II} for parameter values established by the expert
estimation. The triaxial compression test is supposed to be the last
experiment needed to identify parameters, which cannot be identified
by uniaxial or hydrostatic compression tests, i.e. $k_2$ parameter.
Therefore, Young's modulus, Poisson's ratio, $k_1$, $k_3$ and $k_4$
parameters are supposed to be known and their values are taken
directly from~\cite{Caner:2000:MM4II}.

\begin{figure}[!ht]
\centering
\begin{tabular}{cc}
\includegraphics[width=0.45\textwidth,keepaspectratio]{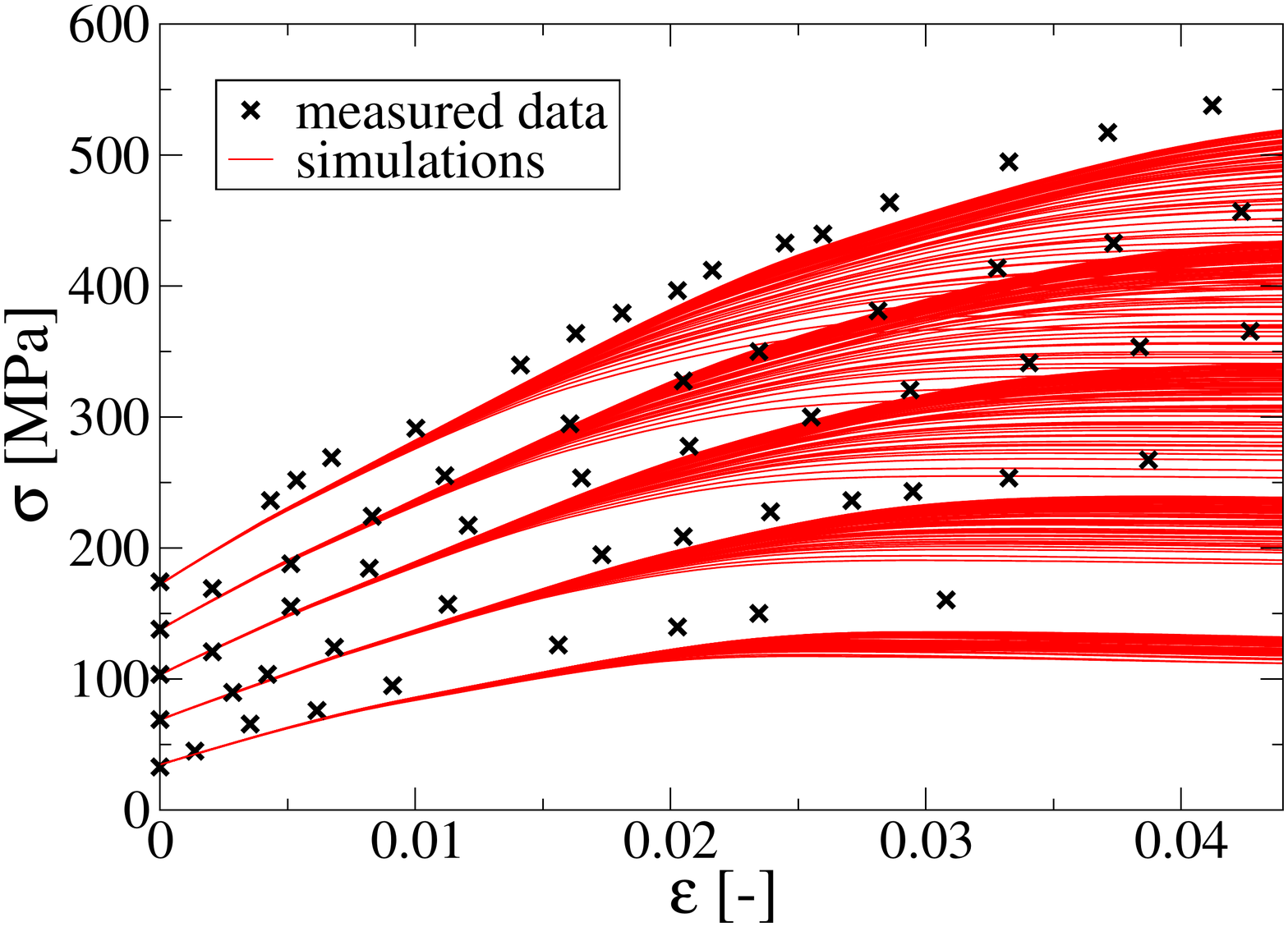} &
\includegraphics[width=0.40\textwidth,keepaspectratio]{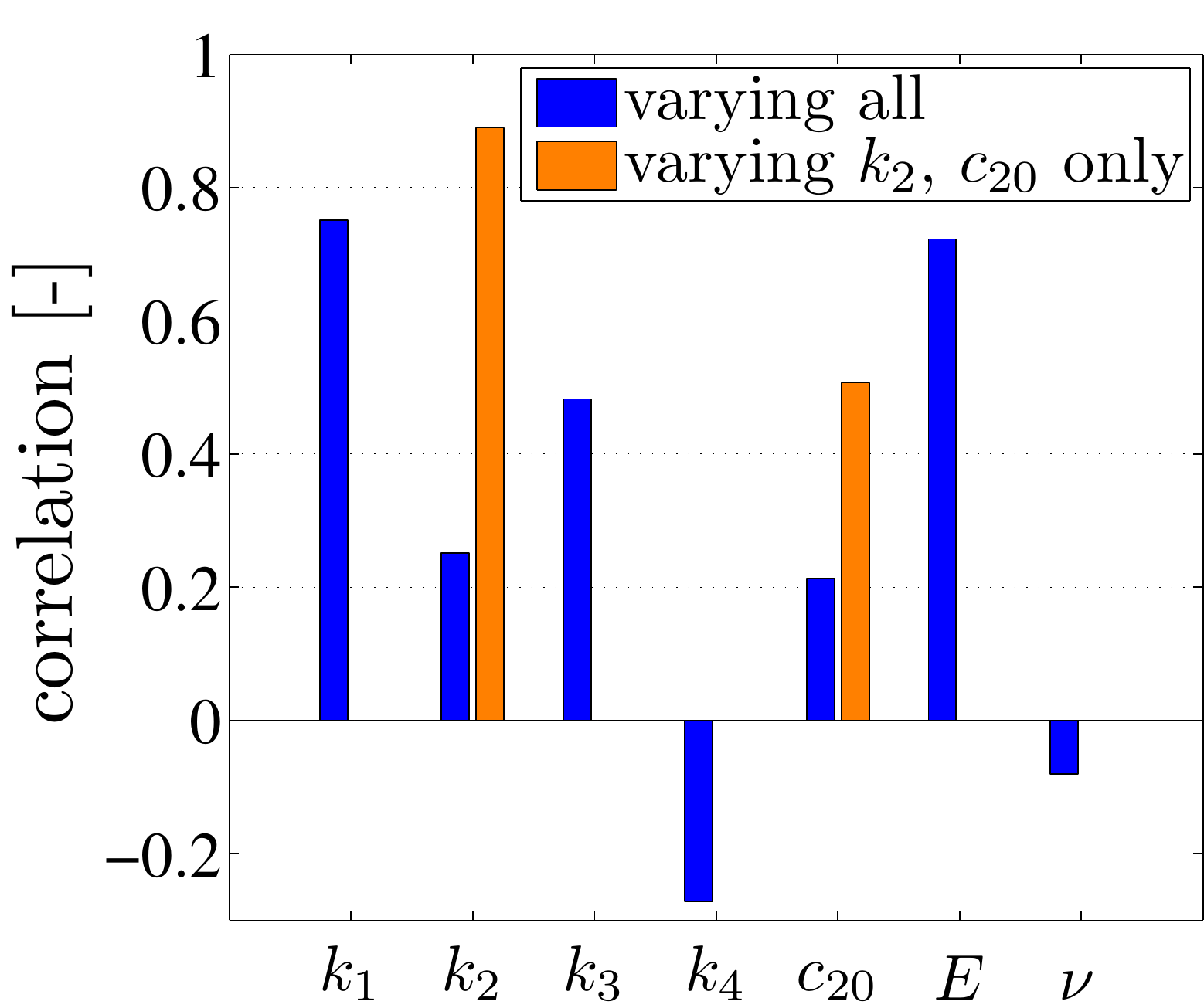} \\
(a) & (b)
\end{tabular}
\caption{Triaxial compression test: (a) comparison of measured data and results of 70 simulations; (b) maximal sensitivities.}
\label{fig_val-tri-bundle1}
\end{figure}

\new{If the triaxial test is the only one available, the other
  parameters can be also identified.  However, they will be predicted
  probably with higher errors, since the sensitivity analysis for
  varying all the parameters shows the highest value of
  sensitivity to Young's modulus only $0.723$ (see Figure
  ~\ref{fig_val-tri-bundle1}b), while its value in uniaxial test
  achieves $0.996$.  Nevertheless, despite the uniaxial or hydrostatic
  tests, the triaxial test is the only one sensitive to all the
  microplane parameters.}

Five different measurements for the triaxial compression test are
available in~\cite{Caner:2000:MM4II} corresponding to five different
levels of a hydrostatic pressure $\sigma_H$ applied to specimens
($\sigma_H \in \{34.5, 68.9, 103.4, 137.9, 172.4\}$ MPa). The bundle
of 70 resulting stress-strain diagrams can be compared with measured
data in Figure~\ref{fig_val-tri-bundle1}a.

In Figure~\ref{fig_val-tri-bundle1}a it is clearly visible that
measured data are remote from all simulated curves. This can be caused
by wrong limits for $k_2$ parameter, i.e. $k_2 \in (100;1000)$,
however an increase of the upper bound by a factor of two did not
bring substantial shift towards experimental
data~\cite{Kucerova:2007:PHD}. The discrepancy can be also caused by
other parameters. Nevertheless, the goal of this Section is not to
identify these parameters from triaxial compression test and
therefore, we will test the ability of a neural network to cope with
this problem.

\begin{table}[!ht]
\centering
\begin{tabular}{c|c|c}
\hline
\bf Parameter & \bf ANN's layout & \bf Input values\\
\hline
$k_2$    & 3 + 2 + 1  & $\sigma_\mathrm{peak}$, $\sigma_{[0.0128]}$, $\sigma_{[0.0308]}$   \\
\hline
\end{tabular}
\caption{Neural network architecture for $k_{2}$ parameter identification.}
\label{tab_tri-ann}
\end{table}

A neural network with an architecture presented in
Table~\ref{tab_tri-ann} was trained on simulated data and then applied
to predict a value of $k_2$ parameter for measured data. Note that
only data from the lowest loading level were used. Errors of ANN's
predictions on training and testing samples are written in
Table~\ref{tab_val-tri-errk2}.
\begin{table}
\centering
\begin{tabular}{c|cc|cc}
\hline
\textbf{Parameter} & \multicolumn{2}{c|}{\textbf{Training data}} & \multicolumn{2}{c}{\textbf{Testing data}} \\
 & Average error & Maximal error & Average error & Maximal error \\
\hline
$k_2$ & 2.63 & 5.82 & 2.37 & 6.23 \\
\hline
\end{tabular}
\caption{Error in ANN's predictions relative to the size of admisible domain given for the $k_2$ parameter in [\%].}
\label{tab_val-tri-errk2}
\end{table}

\begin{figure}[!ht]
\centering
\includegraphics[width=0.8\textwidth,keepaspectratio]{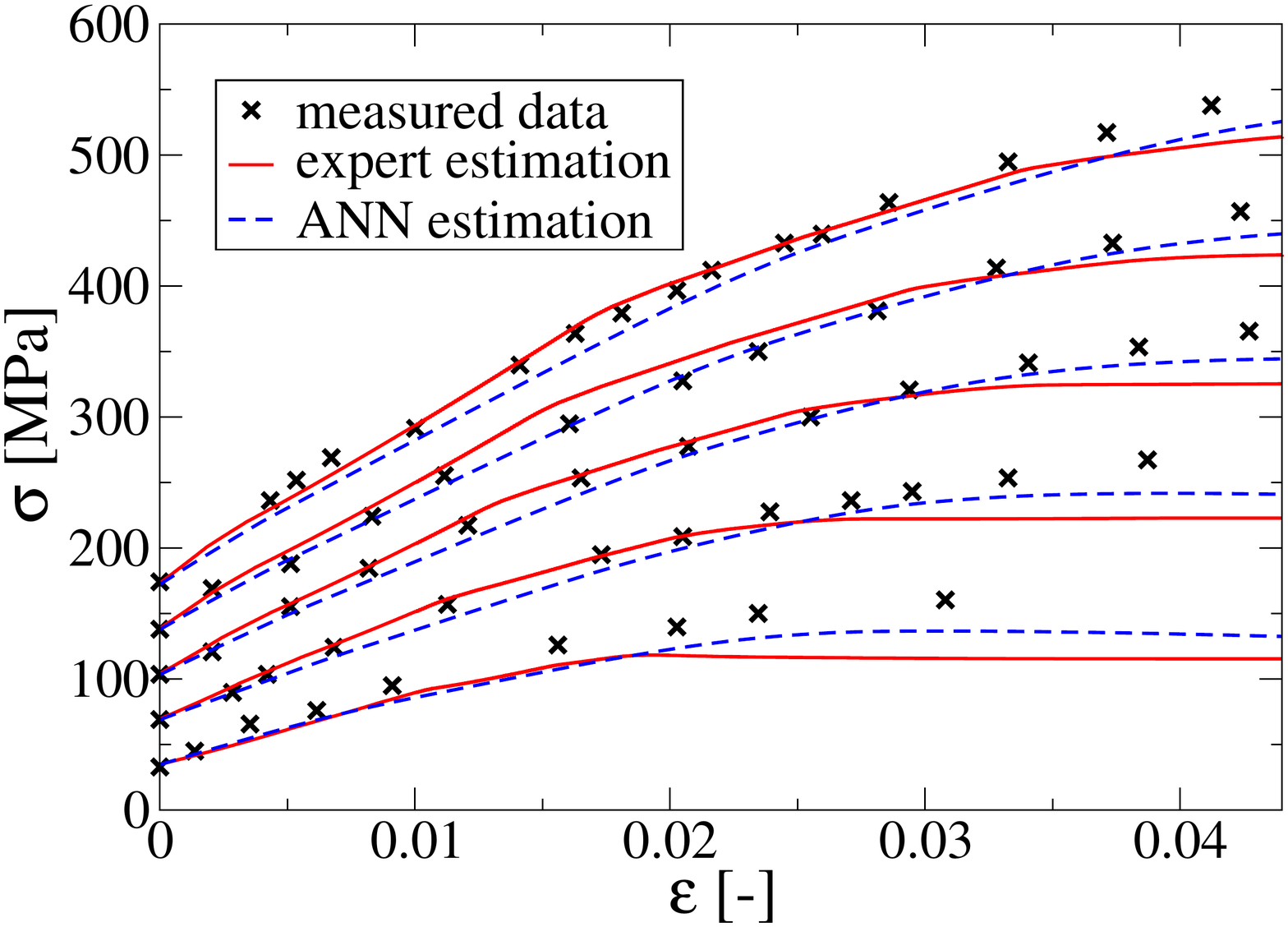}
\caption{Comparison of measured data and simulated diagrams of triaxial compression test for predicted parameters.}
\label{fig_val-tri-comp-pred}
\end{figure}

The prediction of the neural network for measured data is $k_2 =
1193$, while $k_2 = 1000$ is suggested in \cite{Caner:2000:MM4II}. It
is not surprising that the neural network needs to extrapolate and its
prediction exceeds the limit given for $k_2$ parameter. Although
layered neural networks are known to be good in approximations, they
are week in extrapolation. Moreover, there is a question, whether it
is possible to find appropriate value of $k_2$ parameter to fit
measured data, since it is probable that a more important error is
hidden elsewhere. Finally, Figure~\ref{fig_val-tri-comp-pred} shows
the comparison of measured data, a~simulation given in
\cite{Caner:2000:MM4II} and a simulation for the parameter value
predicted by the neural network.

Similarly to previous sections, errors between measured data and
simulated curves can be calculated. Resulting values for all five
experiments corresponding to different levels of hydrostatic pressure
$\sigma_H$ are shown in Table~\ref{tab_val-tri-err}.
\begin{table}[!ht]
\centering
\begin{tabular}{c|cc}
\hline
\textbf{$\sigma_H$ [MPa]} & \textbf{expert estimation \cite{Caner:2000:MM4II}} & \textbf{ANN's predictions} \\
\hline
34.5  & 64 & 43 \\
68.9  & 87 & 60 \\
103.4 & 73 & 45 \\
137.9 & 70 & 39 \\
172.4 & 43 & 60 \\
\hline
\end{tabular}
\caption{Comparison of errors of predicted simulations for triaxial compression test.}
\label{tab_val-tri-err}
\end{table}

Errors in Table~\ref{tab_val-tri-err} are very similar for both
simulations. It is possible to conclude that even for remote input
data the neural network is able to predict a reasonable value of $k_2$
parameter and the resulting simulation seems to be better than the one
obtained by the expert estimation. \new{Moreover, the data from only
  one test with an arbitrary level of hydrostatic pressure $\sigma_H$
  are sufficient for the parameter estimation and prediction of
  material behavior in case of different level of hydrostatic
  pressure.}

%%%%%%%%%%%%%%%%%%%%%%%%%%%%%%%%%%%%%%%%%%%%%%%%%%%%%%%%%%%%%%
%%%%%%%%%%%%%%%%%%%%%%%%%%%%%%%%%%%%%%%%%%%%%%%%%%%%%%%%%%%%%%
\section{Conclusions}
\label{sectconcl}

In this contribution, an example of the engineering problem, which is
difficult to be solved by traditional procedures, was solved using
soft computing methods. Particularly, cascade neural networks were
used to estimate required microplane material model parameters in a
sequential way. As the training procedure, the evolutionary-based
method GRADE extended by CERAF strategy was used. A~number of needed
simulations is reduced by the application of the Latin Hypercube
Sampling method accompanied by the optimization by Simulated
Annealing. The global sensitivity analysis shows not only the influence of
individual parameters but also approximately predicts the errors
produced by neural networks.

\begin{table}[ht]
\centerline{
\begin{tabular}{c|c|c|l}
\hline
\bf Parameter & \bf Test & \bf ANN's topology & \bf ANN's inputs\\
\hline
$E$           & Uniaxial compression & $3 + 2 + 1$ &
$\sigma_{[0.0005]}$, $\sigma_{[0.001]}$, $\sigma_{[0.0015]}$ \\
$k_1$         & Uniaxial compression & $5 + 3 + 1$ &
$\sigma_{[0.0025]}$, $\sigma_{[0.009]}$, $\sigma_{\mathrm{peak}}$, $\epsilon_{\mathrm{peak}}$, prediction of $E$\\
$k_2$         & Triaxial loading     & $3 + 2 + 1$ &
$\sigma_\mathrm{peak}$, $\sigma_{[0.0128]}$, $\sigma_{[0.0308]}$ \\
$k_3$         & Hydrostatic loading  & $5 + 2 + 1$ &
$k_4$, $\epsilon_\mathrm{yield}$, $\epsilon_{[137],\mathrm{load}}$, $\epsilon_{[308],\mathrm{load}}$, $\epsilon_\mathrm{peak}$ \\
$k_4$         & Hydrostatic loading  & $3 + 2 + 1$ &
$k_3$, $\epsilon_\mathrm{peak}$, $\epsilon_{[85.5],\mathrm{unload}}$ \\
$c_{20}$      & Uniaxial compression & $4 + 2 + 1$ &
$\sigma_{[0.003]}$, $\sigma_{[0.004]}$, $\sigma_{[0.006]}$, $\sigma_{[0.008]}$ \\
\hline
\end{tabular}
}
\caption{Final status of M4 identification project}
\label{t:overview}
\end{table}

Results, see Table~\ref{t:overview}, confirm the claims made by
authors~\cite{Caner:2000:MM4II} of the microplane~M4 model on
individual parameters fitting. However, the validation of the complete process cannot be
done, since the experimental data for all three loading tests
performed on one concrete are not available and probably still do not exist.
Therefore, the validation demonstrated in this paper is done only for particular identification
steps.

\begin{figure}[ht!]
\centering
\includegraphics[keepaspectratio,width=0.8\textwidth]{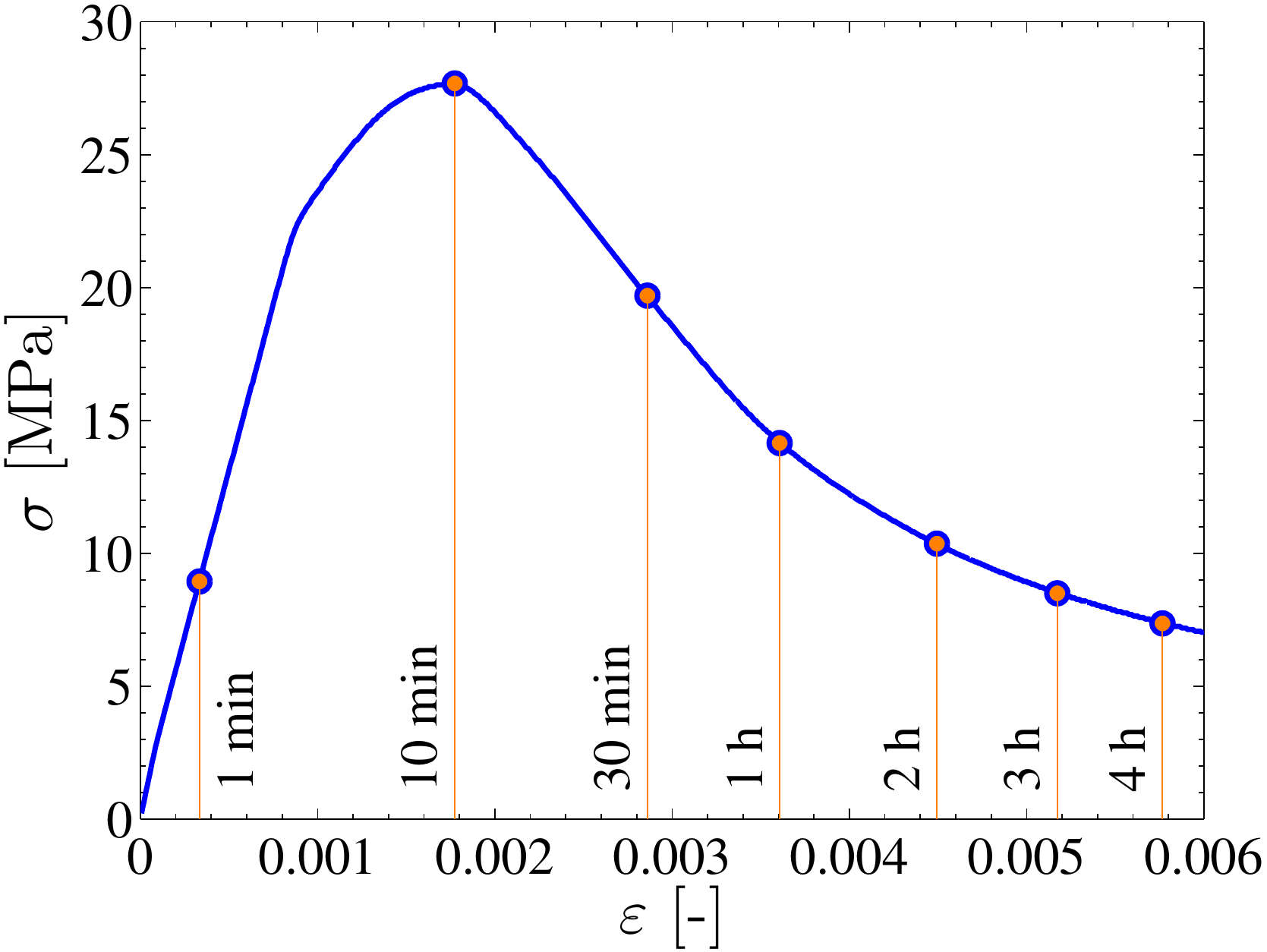}
\caption{Time required to access particular point on the uniaxial load-deflection curve.}
\label{fig_time_evol}
\end{figure}

\begin{figure}[ht!]
\centering
\includegraphics[keepaspectratio,width=0.8\textwidth]{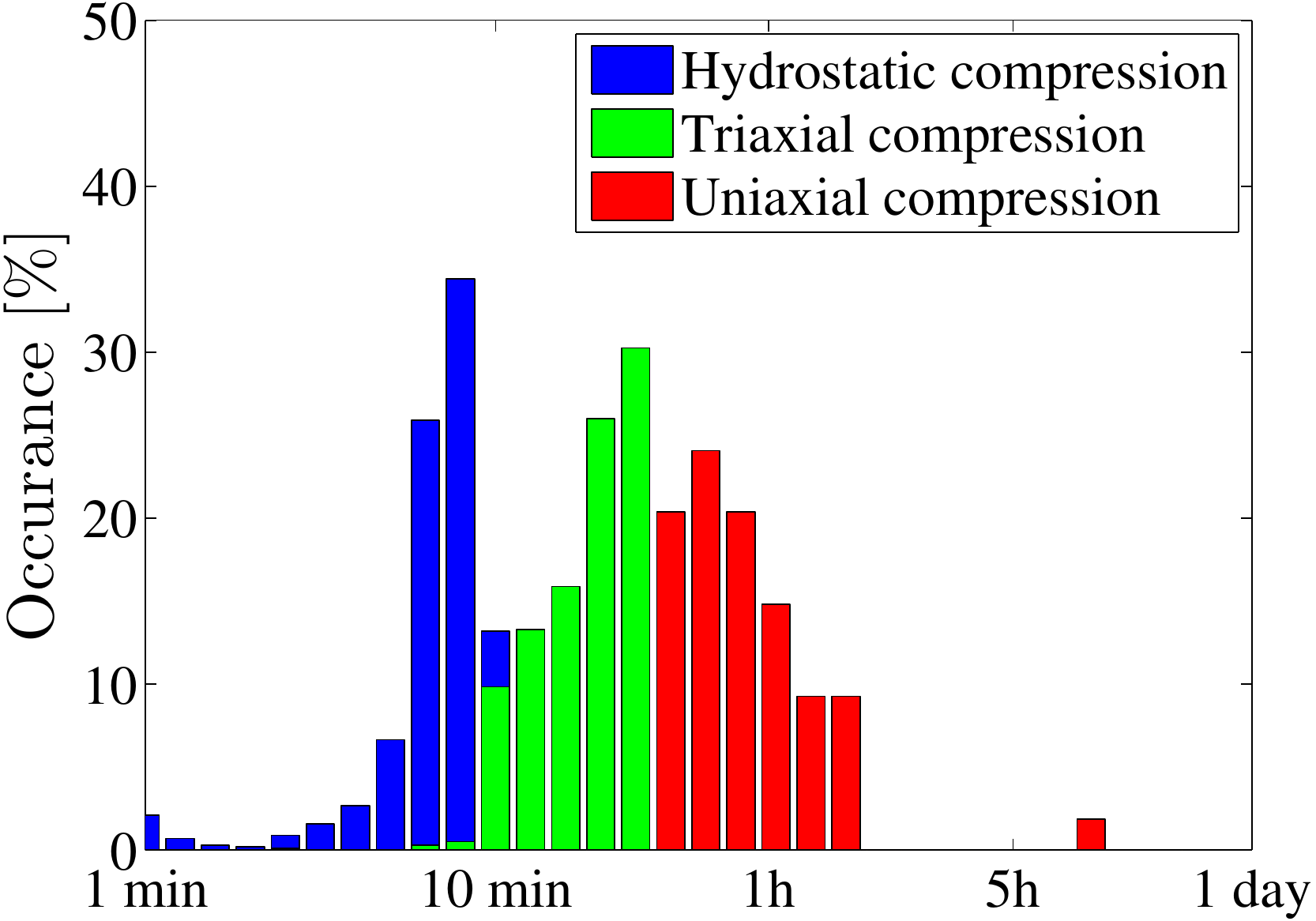}
\caption{Histogram of times needed to run individual tests. The uniaxial test is measured at $\epsilon = 0.004$ to fit the figure.}
\label{fig_time_hist}
\end{figure}

The rather severe disadvantage of the microplane model, and also of
the proposed methodology, is an extreme demand of computational
time\footnote{Note that since the task is symmetric, only 1/8 of the cylinder is actually needed for FEM analysis which shortens the needed time.}. However, especially the uniaxial test dramatically depends on how far the computation will be processed. If we inspect Figure~\ref{fig_time_evol}, every $0.001$ move on the softening branch costs in average one hour on a~single processor PC with the Pentium~IV 3400~MHz processor and 3~GB RAM. However there is a problem that individual compositions lead to big differences in times. This is illustrated in Figure~\ref{fig_time_hist}, where the scatter of times is shown in a form of histograms. The averages are 8~minutes, 19 minutes and one hour for hydrostatic, triaxial and uniaxial compression tests, respectively. However note, that in cases that the material is very soft and the load-deflection curve is long, some computations longer than 6 hours can appear.

Because the identification procedure consists of developing
cascade neural networks, all but one level inverse models should
be recalculated for any new measured data. Fortunately, the most
time consuming simulations of the uniaxial compression test are necessary
for training of the first three neural networks predicting Young's
modulus, Poisson's ratio and $k_1$ and can be used repeatedly for
any new measurement. Then the proposed methodology still needs to
compute 30 uniaxial tests to properly identify $c_{20}$ parameter
and a set of 30 hydrostatic and triaxial tests to fit $k_3$, $k_4$
and $k_2$.

%%%%%%%%%%%%%%%%%%%%%%%%%%%%%%%%%%%%%%%%%%%%%%%%%%%%%%%%%%%%%%%%%%%%%
%%%%%%%%%%%%%%%%%%%%%%%%%%%%%%%%%%%%%%%%%%%%%%%%%%%%%%%%%%%%%%%%%%%%%
%%%%%%%%%%%%%%%%%%%%%%%%%%%%%%%%%%%%%%%%%%%%%%%%%%%%%%%%%%%%%%%%%%%%%
%%%%%%%%%%%%%%%%%%%%%%%%%%%%%%%%%%%%%%%%%%%%%%%%%%%%%%%%%%%%%%%%%%%%%
\vspace*{4mm}
\noindent
{\bf \large Acknowledgement}

\noindent
Financial support for this work was provided by the project CEZ MSM 6840770003 of Ministry of Education, Youth and Sports of the Czech Republic and by the projects P105/11/P370 and P105/12/1146 of the Czech Science Foundation. The financial support is gratefully acknowledged.

%%%%%%%%%%%%%%%%%%%%%%%%%%%%%%%%%%%%%%%%%%%%%%%%%%%%%%%%%%%%%%%%%%%%%
%%%%%%%%%%%%%%%%%%%%%%%%%%%%%%%%%%%%%%%%%%%%%%%%%%%%%%%%%%%%%%%%%%%%%
%%%%%%%%%%%%%%%%%%%%%%%%%%%%%%%%%%%%%%%%%%%%%%%%%%%%%%%%%%%%%%%%%%%%%
%%%%%%%%%%%%%%%%%%%%%%%%%%%%%%%%%%%%%%%%%%%%%%%%%%%%%%%%%%%%%%%%%%%%%
%\bibliography{liter}
%\bibliographystyle{plain}

\end{document}